\documentclass[onecolumn]{aa}
\usepackage{graphicx}

\usepackage{txfonts}

\newcommand{\hcoplus}{\mbox{HCO$^{+}$}}
\newcommand{\hthirteencoplus}{\mbox{H$^{13}$CO$^{+}$}}

\newcommand{\ntwohplus}{\mbox{N$_2$H$^{+}$}}
\newcommand{\htwoco}{\mbox{H$_2$CO}}
\newcommand{\Lsun}{\mbox{$\rm L_\odot\,$}}

\newcommand{\Msun}{\mbox{$\rm M_\odot\,$}}

\newcommand{\microns}{\mbox{$\rm \mu m$}}
\newcommand{\ff}[2]{\mbox{${\rm F}_1 {\rm F}\rightarrow {\rm F}^\prime_1 {\rm F}^\prime = #1\rightarrow #2$}}
\newcommand{\kms}{\mbox{ km\,s$^{-1}$}}
\newcommand{\simlt}{\mbox{\lower.5ex\hbox{$\; \buildrel < \over \sim \;$}}}
\newcommand{\simgt}{\mbox{\lower.5ex\hbox{$\; \buildrel > \over \sim \;$}}}
\newcommand{\cc}{\mbox{cm$^{-3}\,$}}
\renewcommand{\j}[2]{\mbox{J$= #1\rightarrow #2$}}

\usepackage{rotating}
\usepackage{supertabular}
\usepackage{natbib}
\begin{document}
   \title{The Circumstellar Environment of High Mass
   Protostellar Objects. III  Evidence of Infall ?}

   \author{G.~A. Fuller\inst{1}, S.~J. Williams\inst{1}
          \and
          T.K. Sridharan\inst{2}
          }

   \offprints{G.Fuller@manchester.ac.uk}

   \institute{Physics Department, UMIST, P.O. Box 88, Manchester M60
              1QD, UK\\
              \email{Gary.Fuller@manchester.ac.uk,Stewart.Williams@manchester.ac.uk}
         \and
             Harvard-Smithsonian Center for Astrophysics, 60 Garden
              Street, MS 78, Cambridge, MA 02138, USA\\
             \email{tksridha@cfa.harvard.edu}
             }

   \date{Received ; accepted }
   \titlerunning{Infall Towards High Mass Protostellar Objects}
   
   \abstract{The results are presented of a molecular line survey to search
     for the spectral signature of infall towards 77 850$\mu$m continuum
     sources believed to be candidate high mass protostellar objects.  Up to
     six different transitions, \hcoplus\ \j{1}{0}, \j{3}{2}\ and \j{4}{3},
     \htwoco\ $2_{12}-1_{11}$, \ntwohplus\ \j{1}{0} and \hthirteencoplus\ 
     \j{3}{2}, were observed towards each source.  Towards the peak of the
     850$\mu$m emission, \ntwohplus\ was typically strong, with a peak antenna
     temperature of $\sim1.5$K, with a typical linewidth of $\sim2$\kms\ . The
     good agreement between the velocity and velocity width of the \ntwohplus\ 
     and \hthirteencoplus\ emission suggests that both species are tracing
     similar material in the sources.  With respect to the velocity of the
     \ntwohplus, there is a statistically significant excess of blue
     asymmetric line profiles in both the \hcoplus\ \j{1}{0}\ and \htwoco\ 
     transitions. This excess reaches levels similar to that seen towards
     samples of low mass protostars, and suggests that the material around
     these high mass sources is infalling.  We identify 22 promising candidate
     infall sources which show at least one blue asymmetric line profile and
     no red asymmetric profiles. The infall velocity is estimated to be in the
     range of 0.1 \kms\  to 1 \kms\  with an implied mass accretion rate of
     between $2\times10^{-4}$ \Msun/yr and $10^{-3}$ \Msun/yr.

       \keywords{Stars: formation -- ISM: molecules -- ISM: kinematics
         and dynamics -- HII regions } 
}

   \maketitle

\section{Introduction}

The mass, energy and momentum input from high mass stars ($M_*>8$\Msun) 
shapes the physical and chemical properties of the interstellar medium in a
galaxy and can control the ISM's evolution.  Once formed, the high mass stars,
although relatively few in number, rapidly determine a molecular cloud's
future evolution, regulating the formation of lower mass stars. Despite their
importance there is as yet no generally accepted view of how high mass stars
form. Proposed models range from accretion driven formation \citep{mt02,mt03},
similar to that proposed for the formation of low mass stars, competitive
accretion in clusters (Bonnell, Vine \& Bate 2004) or even the agglomeration
of lower mass protostars (Bonnell, Bate \& Zinnecker 1998). 

Recent work identifying candidate high mass protostellar objects (e.g Lumsden
et al.  2002; Molinari et al. 1996; Sridharan et al. 2002) has provided
samples of luminous young sources with which to address some of the
uncertainties about high mass star formation.  Identifying and studying the
accretion flows which collect the material out of which stars form, either
directly or indirectly, is an important aspect of understanding high-mass star
formation.  Evidence for the infall of material around a source can be
provided by analysis of molecular line profiles.  Any ordered motions such as
rotation, outflow or infall can produce recognisable signatures in the line
profiles and their spatial distribution (e.g.  Adelson \& Leung 1988). However
unlike outflow or rotation which give rise to both red and blue asymmetric
lines, infall produces only blue asymmetric lines. The profiles of optically
thick lines from infalling material have stronger blue shifted emission than
red shifted emission (Leung \& Brown 1977; Anglada et al. 1987; Zhou 1992;
Walker, Narayanan \& Boss 1994), provided that the excitation temperature of
the molecules increases towards the centre of the region. This asymmetry
arises as the blue shifted emission from the approaching warm gas on the far
side of the centre of contraction suffers less extinction than the emission
from the receding, nearside, material.  Although for any one object, outflow
or rotation could also produce a blue asymmetric line profile along a
particular line of sight to a source, for an ensemble of sources the presence
of infall should manifest itself as an excess of blue asymmetric line profiles
compared to red asymmetric profiles.  On the other hand, an unbiased sample of
sources dominated by outflows or rotation would be expected to produce roughly
equal numbers of blue and red asymmetric lines.

An excess of blue asymmetric line profiles is now well established towards
low-mass star forming regions, providing strong evidence for infall towards
these regions.  Not only has it been detected in a variety of different
tracers and transitions towards low mass protostars but also towards starless
cores (Mardones et al. 1997; Lee et al.  1999, 2001, 2004; Williams et al
1999; Park et al. 1999; Gregersen \& Evans 2000; Gregersen et al. 2000; Di
Franceco et al. 2001; Belloche et al. 2002). In this paper we present the
first results of a survey for the infall line asymmetry towards a sample of
candidate high mass protostellar objects (HMPOs).



%
%
\begin{table*}[h]
\begin{center}
\caption{Source Positions. The source numbers are from Williams et al. (2004; WFS) 
but  the tabulated positions are the observed positions, which as discussed in
  the text do not necessary precisely agree with those in WFS. The name of the
  corresponding IRAS point source is also given.  }\label{tab:sources}
\begin{tabular}{llcc|llcc}
\hline\hline
Source & IRAS & RA(2000) & Dec(2000) & Source & IRAS &  RA(2000) &
Dec(2000)\\ \hline
2 & 05358+3543 & 05:39:12.6 &+35:45:50   & 62  & 18540+0220 & 18:56:36.7 &+02:24:44 \\  
3 & 05490+2658 & 05:52:11.0 &+27:00:32  & 64  & 18553+0414 & 18:57:53.3 &+04:18:17 \\  
5 & 05490+2658 & 05:52:12.1 &+26:59:35  & 65  & 18566+0408 & 18:59:10.0 &+04:12:13 \\  
6 & 05553+1631 & 05:58:13.5 &+16:31:59   & 66  & 19012+0536 & 19:03:45.1 &+05:40:45 \\  
8 & 18089-1732 & 18:11:51.4 &-17:31:35   & 67  & 19035+0641 & 19:06:01.3 &+06:46:37 \\  
12 & 18090-1832 & 18:12:01.8 &-18:32:01  & 68  & 19074+0752 & 19:09:53.4 &+07:57:14 \\  
13 & 18102-1800 & 18:13:11.5 &-18:00:03  & 70  & 19175+1357 & 19:19:48.4 &+14:02:27 \\ 
14 & 18151-1208 & 18:17:58.2 &-12:07:26  & 71  & 19175+1357 & 19:19:48.8 &+14:02:47 \\ 
15 & 18159-1550 & 18:18:48.3 &-15:48:58  & 72  & 19217+1651 & 19:23:58.7 &+16:57:39 \\  
16 & 18182-1433 & 18:21:09.1 &-14:31:49  & 73  & 19220+1432 & 19:24:19.7 &+14:38:03 \\  
17 & 18223-1243 & 18:25:10.7 &-12:42:28  & 74  & 19266+1745 & 19:28:55.5 &+17:52:00 \\  
18 & 18247-1147 & 18:27:31.5 &-11:45:56  & 75  & 19282+1814 & 19:30:23.1 &+18:20:25 \\  
19 & 18264-1152 & 18:29:14.4 &-11:50:24  & 77  & 19403+2258 & 19:42:27.2 &+23:05:12 \\  
20 & 18272-1217 & 18:30:02.3 &-12:15:38 & 79  & 19410+2336 & 19:43:11.0 &+23:44:06 \\  
21 & 18272-1217 & 18:30:02.7 &-12:15:13 & 80  & 19411+2306 & 19:43:18.1 &+23:13:59 \\  
22 & 18290-0924 & 18:31:43.3 &-09:22:28 & 82  & 19413+2332 & 19:43:29.0 &+23:40:22 \\  
23 & 18290-0924 & 18:31:44.1 &-09:22:16 & 83  & 19471+2641 & 19:49:09.9 &+26:48:52 \\  
25 & 18306-0835 & 18:33:23.3 &-08:33:30  & 85  & 20051+3435 & 20:07:04.5 &+34:44:44 \\  
27 & 18308-0841 & 18:33:33.0 &-08:39:10  & 87  & 20081+2720 & 20:10:13.1 &+27:28:17 \\ 
28 & 18310-0825 & 18:33:47.9 &-08:23:50  & 88  & 20081+2720 & 20:10:15.9 &+27:28:10 \\ 
29 & 18337-0743 & 18:36:27.9 &-07:40:25  & 90  & 20126+4104 & 20:14:25.8 &+41:13:33 \\  
30 & 18345-0641 & 18:37:16.8 &-06:38:32  & 91  & 20205+3948 & 20:22:20.1 &+39:58:20 \\  
33 & 18348-0616 & 18:37:30.3 &-06:14:11  & 93  & 20216+4107 & 20:23:23.7 &+41:17:43 \\  
34 & 18372-0541 & 18:39:55.9 &-05:38:52  & 94  & 20293+3952 & 20:31:12.6 &+40:03:18 \\  
35 & 18385-0512 & 18:41:12.8 &-05:08:59  & 95  & 20319+3958 & 20:33:49.1 &+40:08:35 \\  
36 & 18426-0204 & 18:45:12.2 &-02:01:12  & 96  & 20332+4124 & 20:34:58.8 &+41:34:47 \\ 
37 & 18431-0312 & 18:45:45.6 &-03:09:22  & 97  & 20332+4124 & 20:35:00.9 &+41:35:02 \\ 
38 & 18437-0216 & 18:46:21.6 &-02:12:22 & 99  & 20343+4129 & 20:36:06.4 &+41:39:57 \\ 
39 & 18437-0216 & 18:46:22.4 &-02:14:20 & 100 & 20343+4129 & 20:36:08.0 &+41:39:57 \\ 
42 & 18440-0148 & 18:46:36.4 &-01:45:22  & 101 & 22134+5834 & 22:15:09.6 &+58:49:06 \\  
43 & 18445-0222 & 18:47:10.8 &-02:19:06  & 103 & 22551+6221 & 22:57:06.8 &+62:37:25 \\  
44 & 18447-0229 & 18:47:21.6 &-02:26:11  & 106 & 22570+5912 & 22:58:58.9 &+59:27:42 \\ 
48 & 18449-0158 & 18:47:35.6 &-01:55:26  & 107 & 22570+5912 & 22:59:04.8 &+59:28:19 \\  
50 & 18454-0136 & 18:48:02.1 &-01:33:28  & 108 & 23033+5951 & 23:05:24.7 &+60:08:13 \\  
51 & 18460-0307 & 18:48:39.7 &-03:03:56  & 109 & 23139+5939 & 23:16:10.1 &+59:55:29 \\  
54 & 18470-0044 & 18:49:36.7 &-00:41:05  & 110 & 23151+5912 & 23:17:20.8 &+59:28:49 \\  
55 & 18472-0022 & 18:49:52.2 &-00:18:57  & 111 & 23545+6508 & 23:57:02.4 &+65:24:37 \\ 
57 & 18488+0000 & 18:51:24.8 &+00:04:19  & 112 & 23545+6508 & 23:57:06.0 &+65:24:48 \\ 
59 & 18521+0134 & 18:54:40.6 &+01:38:04  \\
\hline\end{tabular}
\end{center}
\end{table*}

The sources observed were originally selected from the IRAS point source
catalogue by \citet{tk1} based on criteria established by Ramesh \& Sridharan
(1997).  The IRAS sources were selected to be bright with fluxes of greater
than 90Jy and 500 Jy at 60$\mu$m and 100$\mu$m respectively and to have
colours in the IRAS bands which match the \cite{wc} criteria for UCHII regions
but not to have the radio continuum emission expected from a UCHII region.
These characteristics were choosen to identify sources which are luminous and
have dust shells similar to UCHII regions but have not formed an HII region
and so may represent earlier stages in the evolution of mass young stars.
These sources have subsequently been studied by \citet{b1,b2} and Williams,
Fuller \& Sridharan (2004, 2005).

\section{Observations}
\label{sec:obs}

To search for evidence of infall, single point observations were made towards
the peaks of the 850\microns\ SCUBA sources identified by Williams, Fuller \&
Sridharan (2004, hereafter WFS) as associated with the IRAS sources selected
by \citet{tk1}.  The observed positions are listed in Table~\ref{tab:sources}.
Since the target positions were derived from an initial reduction of the SCUBA
images, the positions do not necessarily precisely agree with more accurate
positions published in WFS but typically agree within a few arcseconds.  There
are, however, six sources where the positions are discrepant by 14'' or more
(WFS43, 48, 54, 57, 77, 83).

%
%

\begin{table*}
\caption{Observed transitions and telescope parameters. $\eta$ is the
 telescope beam efficiency, FWHM the full width half maximum beam
 width and $\Delta v$ the channel spacing of the correlator. The
 references for the line frequencies are 1 - \citet{cas95}, 2 -
 \citet{lov92b,lov92a}, 3 - online version of \cite{pp},  4 -
 \citet{ge01}}\label{tab:lines}
  \begin{tabular}[h]{lccccccc}
\hline\hline
Species & Transition & Frequency & Reference & Telescope & $\eta$ &
 FWHM & $\Delta v$ \\ 
 & &  (GHz) & & &  & ($^{\prime\prime}$) & (km/s) \\ \hline
N$_2$H$^+$      & J=$1-0$ & 93.1737767$\pm$0.000007 & 1 & IRAM 30m &0.78 & 26\arcsec & 0.13 \\ 
HCO$^+$         & J=$1-0$ & 89.188518$\pm$0.000009  & 2 & IRAM 30m & 0.78 & 29\arcsec & 0.13 \\
HCO$^+$         & J=$3-2$ & 267.557625 & 3 & IRAM 30m & 0.69  & 20\arcsec & 0.04 \\ 
HCO$^+$         & J=$4-3$ & 356.734288 & 3 & JCMT & 0.63 & 14\arcsec & 0.26 \\ 
H$^{13}$CO$^+$  & J=$3-2$ & 260.255339 $\pm$ 0.0000035  & 4 & JCMT & 0.69 & 20\arcsec & 0.09 \\ 
H$_2$CO         & J$_{kk'}$= $2_{12}-1_{11}$& 140.839518$\pm$0.000007 & 2 &
IRAM 30m & 0.69 & 17\arcsec & 0.08 \\ \hline
  \end{tabular}
\end{table*}



The observations were carried out using the 30m telescope of the
Institut de Radioastronomie Millim\'etric (IRAM)\footnote{IRAM is
  supported by INSU/CNRS (France), MPG (Germany) and IGN (Spain)} near
Granada, Spain and the 15m James Clerk Maxwell Telescope (JCMT) on
Mauna Kea, Hawaii\footnote{The JCMT is operated by the Joint Astronomy
  Centre in Hilo, Hawaii on behalf of the parent organizations
  Particle Physics and Astronomy Research Council in the United
  Kingdom, the National Research Council of Canada and The Netherlands
  Organization for Scientific Research.}. The frequencies of the lines
observed, together with other frequency dependent parameters are
listed in Table~\ref{tab:lines}. The uncertainties in the line
frequencies (where known) introduce a maximum uncertainty of $\pm0.03$
\kms\  (for the \hcoplus\ 1-0) to the derived line velocity.

\begin{table*}
\caption{RMS $T_A^*$ noise levels in spectra.}\label{tab:rms}
\begin{tabular}{lcccc|lcccc|lcccc}
\hline \hline
Source & \multicolumn{3}{c}{HCO$^+$}  & H$_2$CO & Source & \multicolumn{3}{c}{HCO$^+$} & H$_2$CO & Source & \multicolumn{3}{c}{HCO$^+$}  & H$_2$CO \\ \cline{2-4}\cline{7-9} \cline{12-14}
& 1-0 & 3-2 & 4-3 & &  &  1-0 & 3-2 & 4-3 &  &  &  1-0 & 3-2 & 4-3 & \\ 
& (K)& (K)& (K) & (K) &  &  (K)& (K)& (K) & (K)& &  (K)& (K)& (K) & (K) \\ 
\hline 
2 &  0.07 & \ldots &  0.10 & \ldots &              36 &  0.11 & \ldots &   0.04 &  0.08 &         77 &  0.06 & \ldots &   0.08 &  0.07        \\
3 &  0.06 & \ldots &   0.13 & \ldots &                   37 &  0.10 & \ldots &   0.05 &  0.07 &         79 &  0.03 & \ldots &   0.09 &  0.07        \\
5 &  0.06 & \ldots & \ldots &  \ldots &              39 &  0.04 & \ldots &   0.04 &  0.09 &         80 &  0.06 & \ldots &   0.59 &  0.08        \\
6 &  0.07 & \ldots &   0.10 & \ldots &              38 &  0.07 &  0.32 &  0.03 &  0.10 &    82 &  0.04 & \ldots &   0.08 &  0.08        \\
8 &  0.10 &  0.42 &  0.11 &  0.08 &   42 &  0.07 & \ldots &   0.03 &  0.08 &         83 &  0.06 & \ldots &   0.09 &  0.05        \\
12 &  0.07 &  0.42 &  0.07 &  0.09 &  43 &  0.07 &  0.20 &  0.08 &  0.06 &    85 &  0.05 & \ldots &   0.16 &  0.07        \\
13 &  0.04 & \ldots &   0.09 &  0.07 &       44 &  0.08 & \ldots &   0.05 &  0.08 &         88 &  0.05 & \ldots &   0.14 &  0.07        \\
14 &  0.02 & \ldots &   0.08 &  0.06 &       48 &  0.08 &  0.27 &  0.07 &  0.08 &    87 &  0.05 & \ldots &   0.08 &  0.08        \\
15 &  0.06 & \ldots &   0.07 &  0.06 &       50 &  0.06 &  0.23 &  0.05 &  0.05 &    90 &  0.06 &  \ldots &  0.11 &  0.04   \\
16 &  0.04 & \ldots &   0.07 &  0.06 &       51 &  0.08 & \ldots &   0.05 &  0.07 &         91 &  0.05 & \ldots &   0.11 &  0.07        \\
17 &  0.04 & \ldots &   0.12 &  0.06 &       54 &  0.07 &  0.26 &  0.08 &  0.07 &    93 &  0.04 &  0.23 &  0.09 &  0.06   \\
18 &  0.18 & \ldots &   0.18 &  0.06 &       55 &  0.08 & \ldots &   0.06 &  0.07 &         94 &  0.06 &  \ldots &  0.09 &  0.07   \\
19 &  0.06 & \ldots &   0.11 &  0.06 &       59 &  0.09 &  0.28 &  0.05 &  0.07 &    95 &  0.05 &  0.23 &  0.06 &  0.06   \\
21 &  0.09 & \ldots &   0.17 &  0.06 &       62 &  0.06 &  0.29 &  0.04 &  0.07 &    96 &  0.07 &  0.23 &  0.07 &  0.07   \\
20 &  0.05 &  0.39 &  0.13 &  0.09 &  64 &  0.03 &  0.19 &  0.05 &  0.05 &    97 &  0.07 &  0.20 &  0.08 &  0.06   \\
23 &  0.09 & \ldots &   0.09 &  0.07 &       65 &  0.07 &  0.23 &  0.26 &  0.05 &    99 &  0.05 &  0.21 &  0.15 &  0.07   \\
22 &  0.07 &  0.34 &  0.08 &  0.09 &  66 &  0.04 &  0.24 &  0.07 &  0.07 &    100 &  0.05 &  0.20 &  0.09 &  0.07  \\
25 &  0.10 & \ldots &   0.17 &  0.08 &       67 &  0.05 &  0.24 &  0.09 &  0.07 &    101 &  0.09 &  0.28 &  0.10 &  0.07  \\
27 &  0.09 & \ldots &   0.12 &  0.08 &       68 &  0.06 &  0.23 &  0.05 &  0.06 &    106 &  0.07 &  0.32 &  0.07 &  0.07  \\
28 &  0.09 & \ldots &   0.09 &  0.07 &       71 &  0.08 & \ldots &   0.10 &  0.07 &         107 &  0.06 &  0.25 &  0.08 &  0.06  \\
29 &  0.09 & \ldots &   0.10 &  0.08 &       70 &  0.07 &  0.24 &  0.08 &  0.06 &    108 &  0.07 &  0.25 &  0.06 &  0.06  \\
30 &  0.09 & \ldots &   0.07 &  0.08 &       72 &  0.05 & \ldots &   0.10 &  0.08 &         109 &  0.07 &  0.16 &  0.07 &  0.05  \\
33 &  0.08 & \ldots &   0.38 &  0.08 &       73 &  0.07 &  0.24 &  0.11 &  0.07 &    110 &  0.05 &  0.31 &  0.05 &  0.07  \\
34 &  0.09 & \ldots &   0.04 &  0.07 &       74 &  0.07 & \ldots & \ldots &   0.08 &              111 &  0.07 &  0.16 &  0.08 &  0.04  \\
35 &  0.10 & \ldots &   0.08 &  0.07 &       75 &  0.06 & \ldots &   0.09 &  0.08 &         112 &  0.07 &  0.25 &  0.22 &  0.07  \\
\hline
\end{tabular}
\end{table*}


At the JCMT the \j{4}{3}\ transition of \hcoplus\ and the \j{3}{2}\ transition
of \hthirteencoplus\ were observed using the RxB3 and RxA3 receivers
respectively over the periods from April to June 2001 and May to June
2002. The \hcoplus\ observations used both polarization channels of RxB3 and
after checking that the two channels agreed in line intensity and center
velocity the two channels were averaged together for analysis. A portion of
these observations were made using in-band frequency switching. This resulted
in curved, but smooth baselines which were removed using polynomial fits. The
\hthirteencoplus\ observations were position switched using the single channel
of RxA3.  The off positions were typically 14' away from the sources at a
position (+10',+10') from the source.  The typical rms noise level reached
during the observations was 0.08K for the \hcoplus\ and 0.1 K for the
\hthirteencoplus\ and the typical system temperatures were around 750K and
390K respectively.  All the JCMT observations made use of the DAS
autocorrelator spectrometer. The channel separation used for each of the
transitions is listed in Table~\ref{tab:lines}. The JCMT data were all
initially reduced using the JCMT SPECX package
and then imported to the IRAM spectral line reduction package CLASS
for comparison with the IRAM observations.

The IRAM observations were carried out between December 2001 and January 2002
and in August 2002. The \ntwohplus\ and \hcoplus\ \j{1}{0} transitions were
simultaneously observed in different polarizations using the two 3mm
receivers. The typical system temperatures and rms noise levels in the
resulting spectra were 120K and 0.07 K, and 130K and 0.07K for the \hcoplus\
and \ntwohplus\ respectively.  The \htwoco\ line was then observed separately
using both the facility 2mm receivers. The two \htwoco\ spectra were checked
for consistency before being averaged together for analysis resulting in
spectra with typical system temperatures and noise levels of $\sim190$K and
$\sim0.07$K respectively.  During the August 2002 observations the \j{3}{2}\
transitions of \hcoplus\ and \hthirteencoplus\ were observed, one with each of
the 1mm receiver systems, simultaneously with \htwoco. In no case did the
observations of \hthirteencoplus\ provide a higher signal to noise detection
than the JCMT observations and only the JCMT data on this transition are
presented here and considered in the analysis. The rms noise levels for each
of the HCO$^+$ and \htwoco\ spectra are given in Table~\ref{tab:rms}.  All the
IRAM observations were made using position switching with the same off
positions as used at the JCMT.  The sources were observed using the
autocorrelator spectrometer and during the August 2002 run, the newer VESPA
correlator. The IRAM data were reduced using the CLASS package.

Considering both the JCMT and IRAM data, there was only evidence of
off-position emission contaminating the observations in about four
spectra in total and in all cases the emission at the off position was
considerably weaker than towards the source, easily identifiable and
did not affect the determination of the line peak velocity.  The
pointing and focus were regularly checked during the observations with
both telescopes. The largest pointing errors seen were $\sim3''$ and
typically the pointing offsets were less than 2''.

Spectra towards the sources are shown in Figure~1
and
2. The hyperfine structure of the \ntwohplus\ 
transition produces three components in the spectra. None of the other
transitions have hyperfine structure.

\subsection{Reduction}\label{sec:red}

The reduction and analysis of the line profiles to look for evidence of infall
asymmetry is primarily concerned with the line velocities and widths, and we
will limit the discussion to just these aspects of the emission. The line
intensities and column densities which can be infered from the optically thin
\ntwohplus\ and \hthirteencoplus\ emission can provide useful information on
the circumstellar material, but we will delay the discussion of these aspects
of the emission until a later work.

  \includegraphics{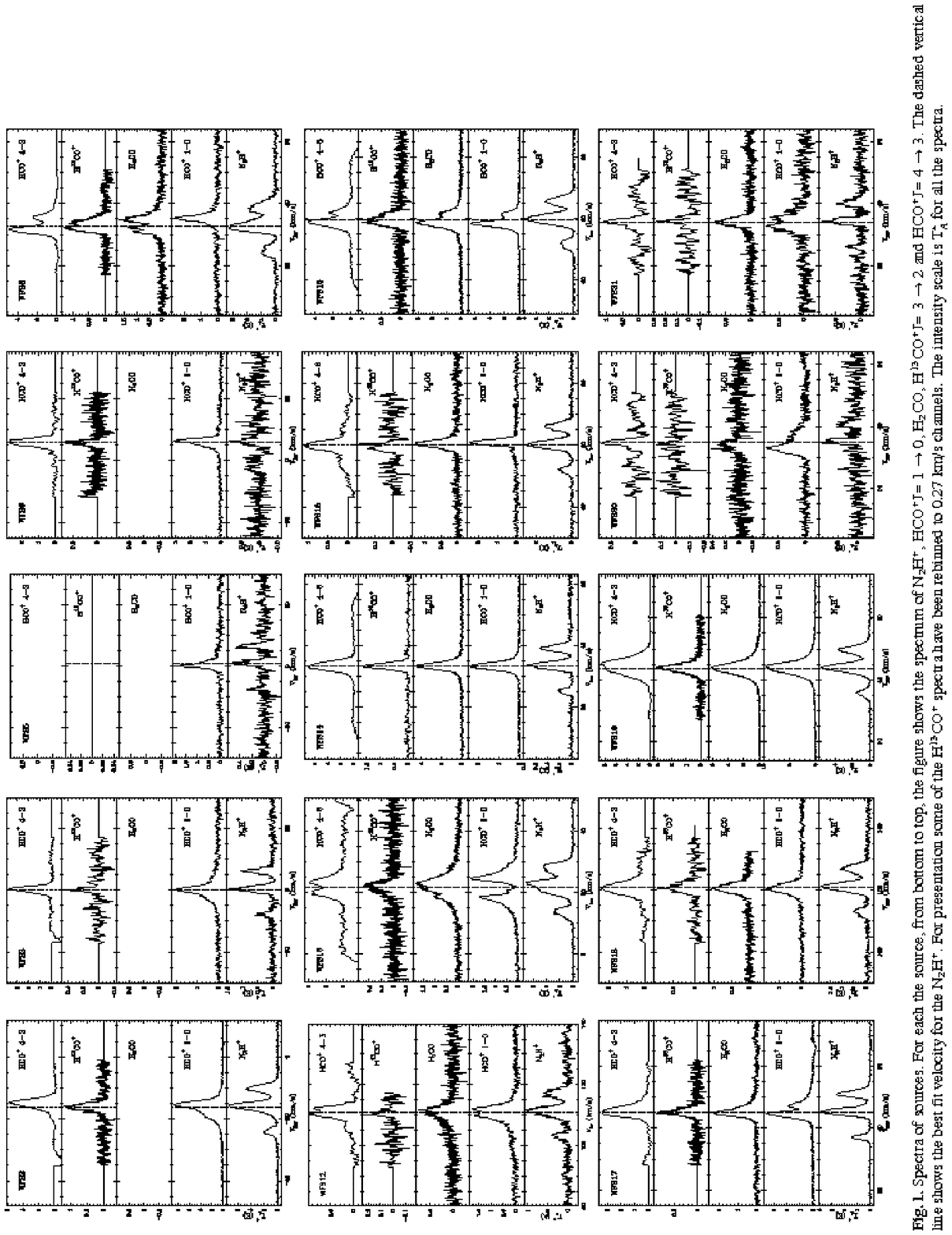}  
\newpage
  \includegraphics{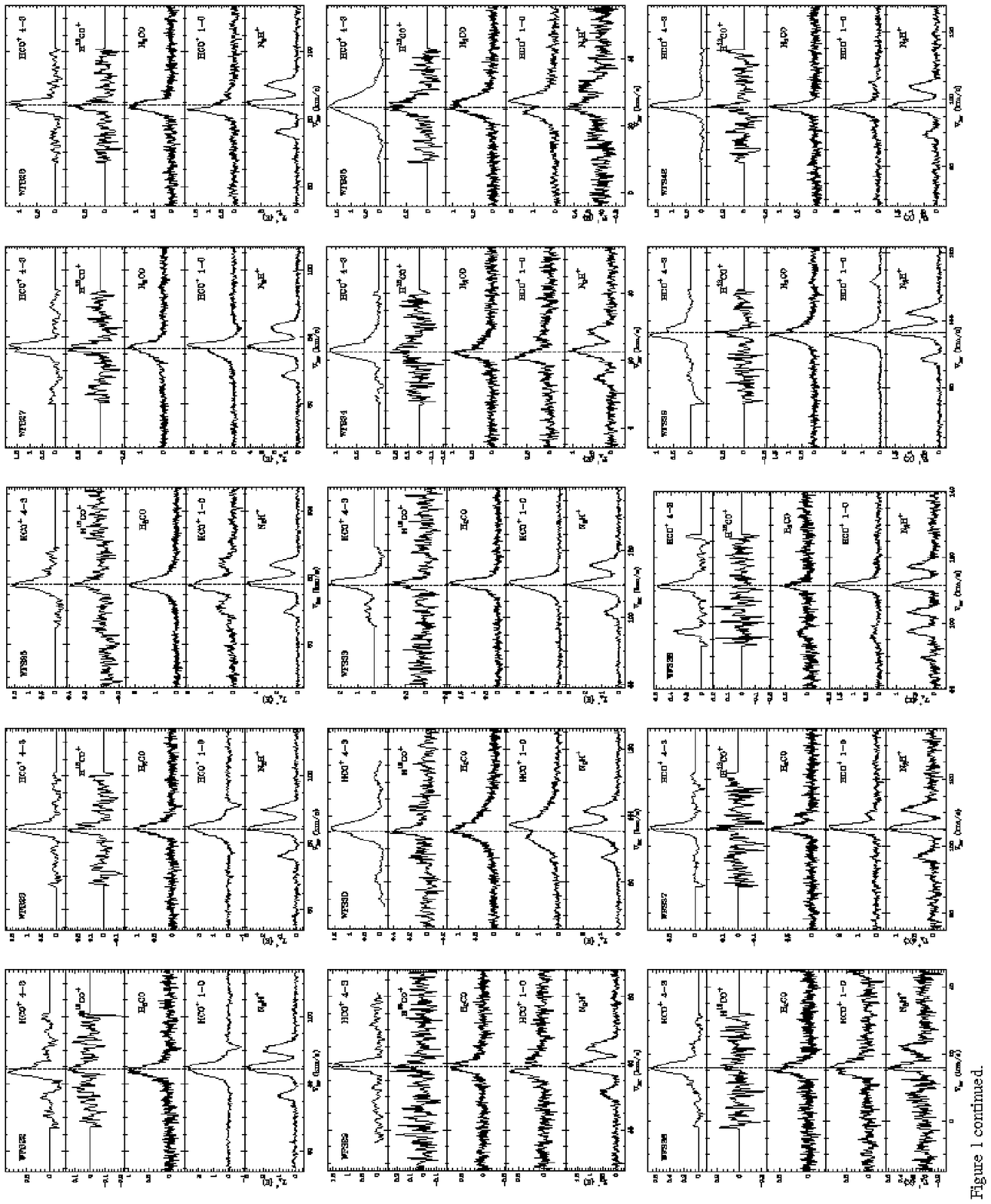}  
\newpage
  \includegraphics{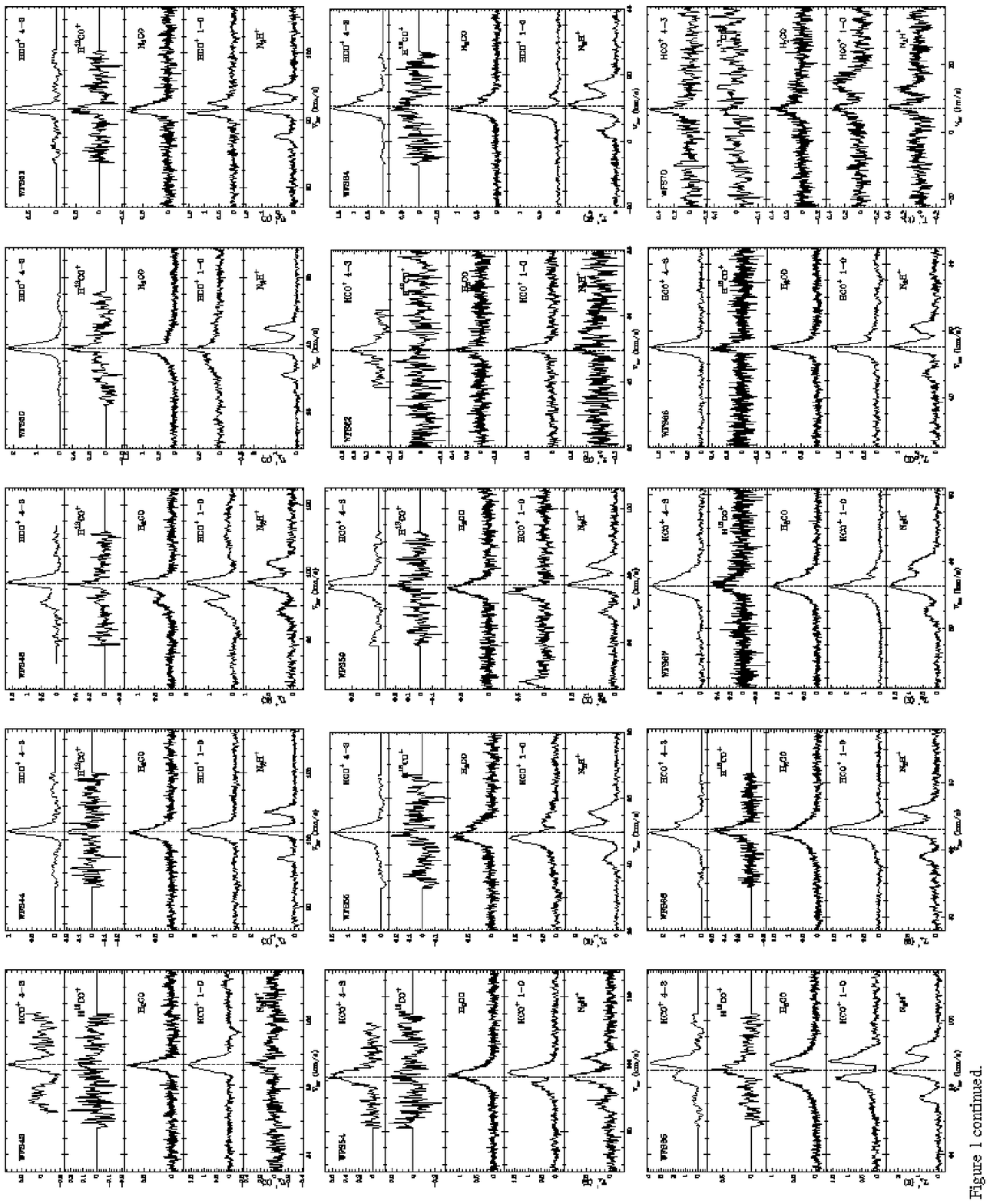}  
\newpage
  \includegraphics{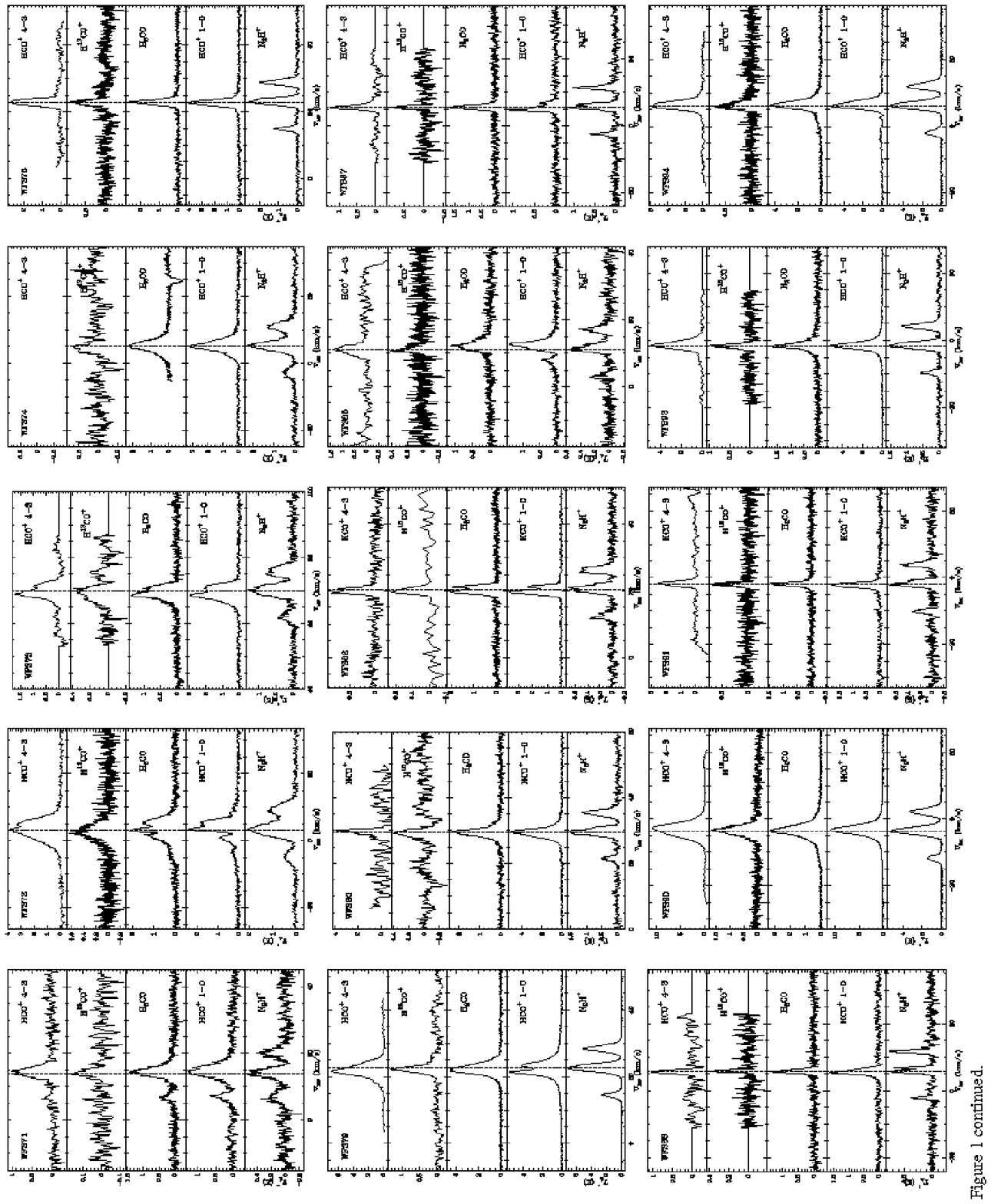}  
\newpage
  \includegraphics{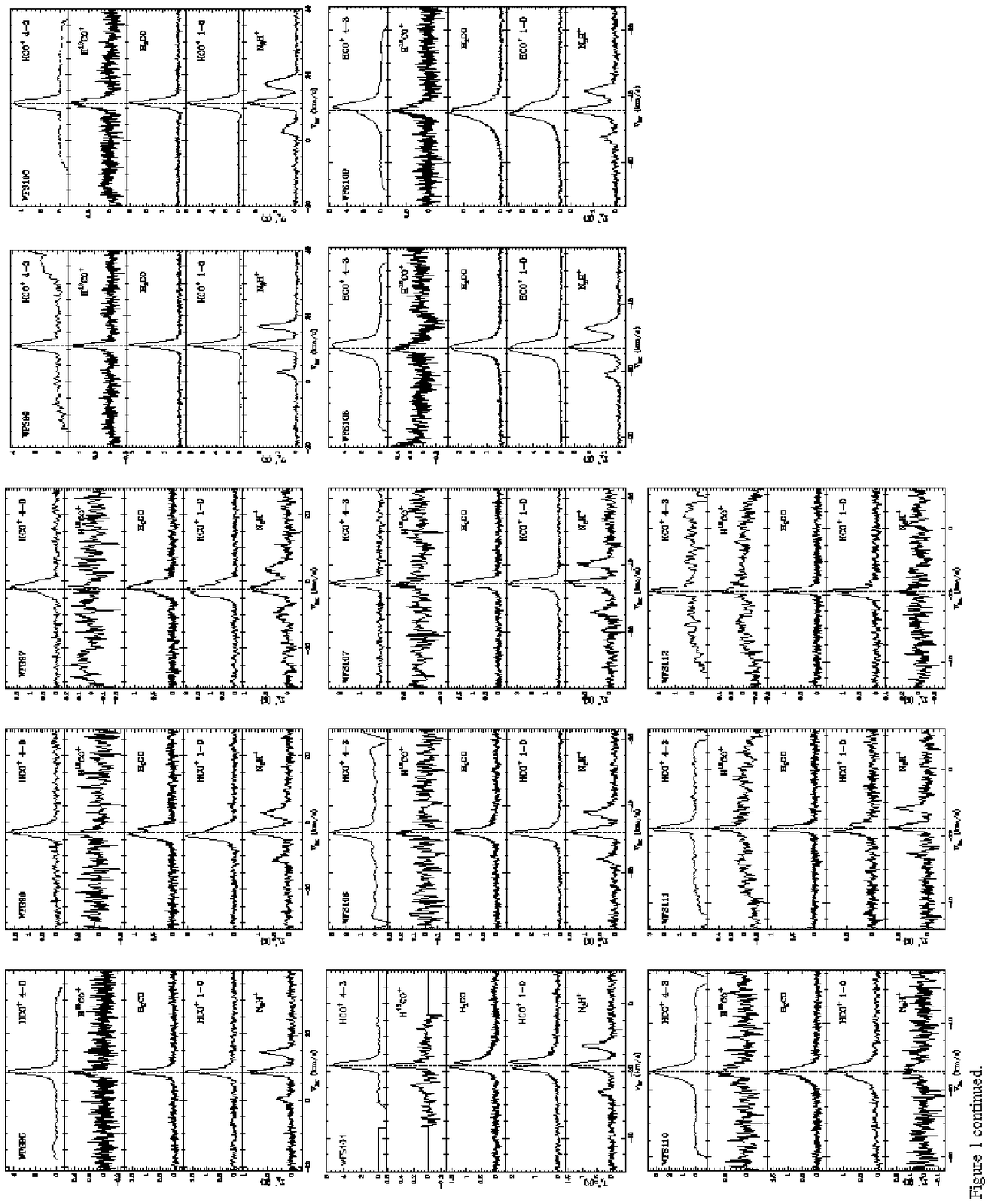}  

The \ntwohplus\ spectra were fitted with the seven hyperfine components given
in Table~\ref{tab:hfs} using the hyperfine fitting routines in CLASS to
determine the properties of the emission.  The parameters defining the
hyperfine components are listed in Table~\ref{tab:hfs} and the results for the
fit for each source are given in Table~\ref{tab:n2hpfits}.  The majority of
the spectra are well fitted by a single gaussian velocity component, however
for some sources, for example WFS2, a single component is not a good model.
In Table~\ref{tab:n2hpfits} these sources can be identified as the sources for
which the residual in the region where the line was fitted is significantly
larger than in a region of the spectrum free from emission. Most of the
spectra are also consistent with optically thin emission, although again
there are exceptions, for example WFS~13 and WFS~38.


  \includegraphics[height=21cm]{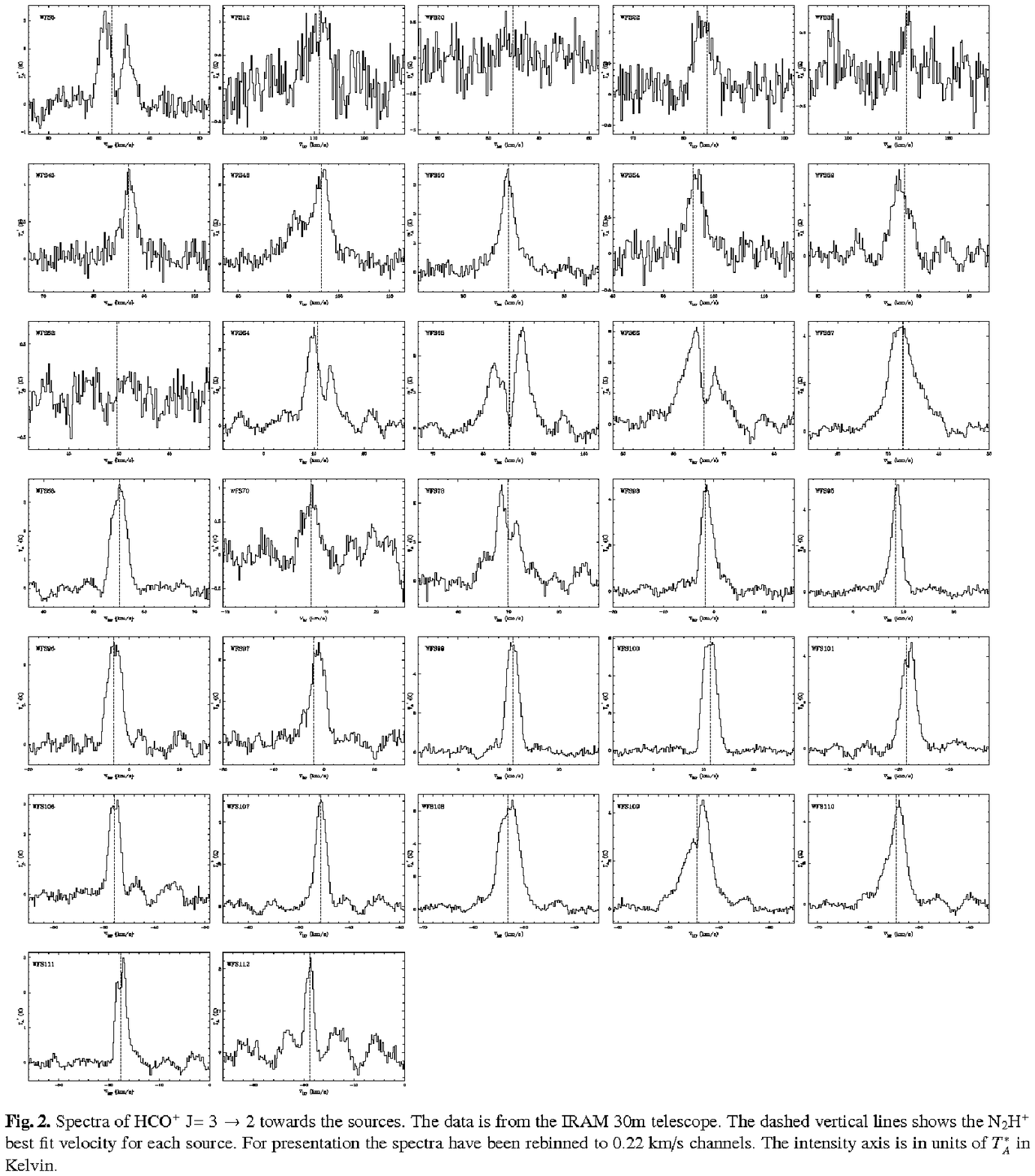}  
\setcounter{figure}{2}

The only sources towards which \ntwohplus\ was not detected are WFS~57,
WFS~83, and WFS~103. 
Towards WFS77
the \ntwohplus\ line was very weak and only very poorly detected.  No
\hthirteencoplus\ emission was detected towards these sources and these
sources are not considered any further in the analysis below.
The \ntwohplus\ line is also very weak towards WFS~110 and WFS~112
and these sources are  also excluded from the analysis below.  The
source WFS~48,
whilst observed here and by \citet{wfs1}, is not part of the \citet{tk1}
sample and is not included in any of the analysis discussed below. For
consistency in probing the material along the line of sight to the central
sources in each object, the two remaining objects which were observed at
(relatively) large offsets from the submillimetre peak (WFS43 and 54; Sec 2)
were also excluded from the analysis below.

%
%

\begin{table}[tb]
\caption{Hyperfine structure of \ntwohplus. The velocity offset is
  measured with respect to the \ff{2 3}{1 2} component for which we
  adopt a frequency of  93.1737767 GHz. The statistical weight of the
  hyperfine component is given in the column labelled Weight.}\label{tab:hfs}
\begin{center}
  \begin{tabular}[h]{lcc}\hline\hline
Component & Velocity offset & Weight\\ 
\mbox{${\rm F}_1 {\rm F}\rightarrow {\rm F}^\prime_1 {\rm F}^\prime$}&   (km/s) & \\ \hline
$1 0\rightarrow 1 1$  &  6.94  &       0.04\\
$1 2\rightarrow 1 2$  &  5.98  &       0.18\\
$1 1\rightarrow 1 0$  &  5.54  &       0.11\\
$2 2\rightarrow 1 1$  &  0.96  &       0.18\\
$2 3\rightarrow 1 2$ &  0.00  &       0.26\\
$2 1\rightarrow 1 1$  & -0.61  &       0.11\\
$0 1\rightarrow 1 2$  & -8.01  &       0.11\\\hline
   \end{tabular}
\end{center}
\end{table}


The results of fitting single Gaussian components to the
\hthirteencoplus\ lines for which the line peak intensity is greater
than 2.5 times the rms noise are given in Table~\ref{tab:h13cofits}.
Comparing the rms of the residuals of the line fit with the rms
noise in an emission free region of the spectrum shows that all these
spectra are well modelled by these single Gaussian components.  The 2.5
times the rms noise limit corresponds to a detection in the integrated
intensity of more than 10 for a line with a typical width of 2\kms.

The other transitions show a range of line shapes ranging from Gaussian
profiles to profiles with multiple peaks and line wings. To measure the line
peak velocity of these lines, a parabolic fit over a number of channels around
the line centre, covering the velocity range out to where the line reached
about half of its peak intensity, was used. This method was preferred to using
Gaussian fits to avoid any line wings influencing the fitted velocities. These
parabolic fits were all inspected to confirm that they provided a good measure
of the velocity of the line peak. The results of these parabolic fits are
shown in Table~\ref{tab:asymm}.  As discussed below (\S~\ref{sec:asymm}),
since the \hcoplus\ \j{4}{3}\ transition shows the least range in line shape
with the majority of lines being Gaussian, these lines were also fitted with
single Gaussian components for comparison with the parabolic fits. The results
for the line velocity from both the parabolic and Gaussian fits are also
listed in Table~\ref{tab:asymm}.

\section{Velocity Structure}

Since multiply peaked line profiles and line shoulders can be due to the
presence of multiple, unrelated velocity components along the line of sight,
rather than the self absorption expected in the presence of infall, it is
important to identify the systemic velocity of the gas in which the sources
are forming.  For this purpose an optically thin spectral line is needed as
the line peak velocity of such a line will reflect the mass-weighted average
velocity along the line of sight.  In this study we have two species which can
serve this purpose: \ntwohplus\ and \hthirteencoplus.

For the \ntwohplus\ emission the intensity ratio of the hyperfine components
of the transition provide a direct measure of the line optical depth. The
optical depth derived from the hyperfine fits (Table~\ref{tab:n2hpfits}) show
that for the majority of sources the line has an optical depth of less than
1. Although there are some sources with larger values, even for these sources
it is unlikely that optical depth significantly affects the estimate of
the systemic velocity as the optical depth at the line peak is at very most
0.4 of this total value for broad lines and as little as 0.25 for narrrow
lines.  Although it is harder to estimate the \hthirteencoplus\ optical depth,
the weakness of the line compared to the \hcoplus\ 1-0 emission and its simple
Gaussian line profile, consistent in width with the \ntwohplus
(\S~\ref{sec:v}), suggest the emission is optically thin.

\subsection{Systemic Velocities}
\label{sec:v}

\begin{figure*}
  \centering
  \includegraphics[angle=-90, width=15cm]{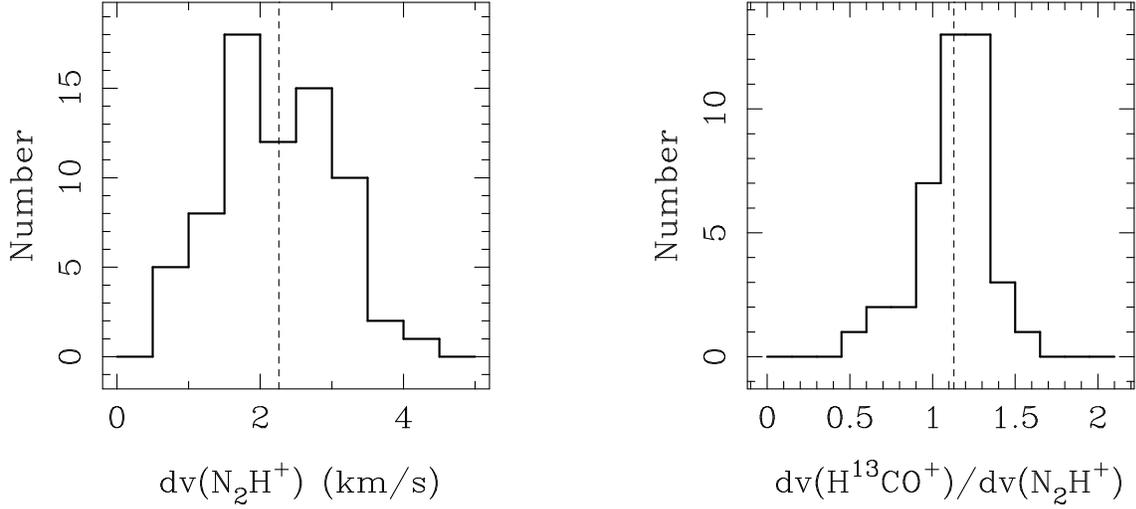}  
  \caption{The left panel shows the distribution of \ntwohplus\ linewidths of
    the sources. The vertical dashed line indicates the median \ntwohplus\ 
    linewidth of 2.26 \kms\ . The right panel shows the ratio of
    \hthirteencoplus\ linewidth to \ntwohplus\ linewidth.  The median value of
    1.13 indicated by the dashed vertical line. }
  \label{fig:comparedv}
\end{figure*}

\begin{figure}
  \centering
  \includegraphics[angle=-90, width=6cm]{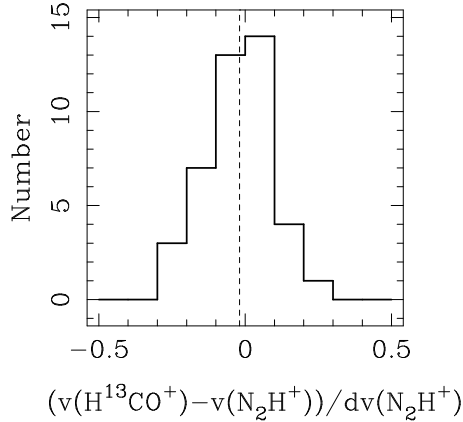}
  \caption{Comparison of \ntwohplus\ and \hthirteencoplus\
    velocities. The figure shows the difference in velocity between
    \hthirteencoplus\ and \ntwohplus\ in units of the \ntwohplus\ linewidth.
    The vertical dashed line marks the median value of -0.02.  All the sources
    for which the \hthirteencoplus\ was detected with a line peak temperature
    greater than 2.5 times the rms noise level in the spectrum are included in
    the figure.}  \label{fig:comparev}
\end{figure}

Since we have observations of two tracers which different authors have
used to define source systemic velocities in searches for infall
\citep{mar,ge00}, we can compare the properties of these two
lines to determine whether these species are tracing the same
material.  One test is to compare the linewidth of the
\hthirteencoplus\ and that derived from the hyperfine fit to the
\ntwohplus.  Figure~\ref{fig:comparedv} shows the \ntwohplus\ 
linewidths and the ratio of the \hthirteencoplus\ linewidth to the
\ntwohplus\ linewidth.  The mean linewidth ratio is $1.11\pm0.03$ (the
quoted uncertainty is the standard error on the mean). Although this
result suggests that on average the linewidth of the \hthirteencoplus\ 
is $\sim10$\% larger than that of the \ntwohplus, this relatively
small difference suggests that these two species trace very similar
material.  Figure~\ref{fig:comparev} shows the difference in line
velocity in units of the \ntwohplus\ line width.  The mean value of
this quantity of $-0.03$, with a standard error on the mean of $0.02$,
again showing that in general these lines are tracing the same
material associated with the forming stars.

Since these tracers appear to be tracing the same material around the
sources, but our observations of \ntwohplus\ have higher signal to
noise and we detect it towards a larger fraction of the sources, we
choose to adopt the \ntwohplus\ velocity as the source systemic
velocity.

\begin{sidewaystable}
\begin{center}
\caption{Hyperfine fits to \ntwohplus\ spectra. For each source the
  table gives (in order) the velocity of the emission, the FWHM velocity width
  of the emission, the line optical depth, the line peak temperature in units
  of $T_A^*$, derived from the fit to the hyperfine structure, and the rms in
  an emission-free region of the spectrum and the rms residual in the region
  of the line after fitting to the hyperfine structure of the line. The values
  in parenthesis are the one standard deviation uncertainties in the
  associated quantities. The quoted optical depth is the peak optical depth
  the emission would have in the absence of the hyperfine
  structure.}\label{tab:n2hpfits} {\small
\begin{tabular}{ccccccc|ccccccc}
\hline\hline
Source & V & dv & $\tau$ & $T_{peak}$ & Baseline & Line & Source & V & dv & Tau & $T_{peak}$ & Baseline & Line \\
               &  &    &     &      & rms  & rms & &  &    &     &      & rms  & rms \\
               & (km/s) & (km/s) & & (K) & (K) & (K) & & (km/s) &
               (km/s) & & (K) & (K) & (K) \\
\hline
2 & -16.10 (  0.01) &   2.71 (  0.01) &   0.11 (  0.01) &   7.03  &   0.07 &   0.26 & 55 &  49.65 (  0.01) &   3.17 (  0.02) &   0.10 (  0.02) &   2.26  &   0.07 &   0.08 \\ 
 5 &   0.72 (  0.03) &   1.00 (  0.08) &   1.09 (  1.25) &   0.48  &   0.07 &   0.07 & 59 &  77.18 (  0.02) &   2.54 (  0.04) &   0.11 (  0.01) &   1.56  &   0.07 &   0.10 \\ 
 3 &   0.22 (  0.01) &   1.43 (  0.05) &   0.47 (  0.38) &   1.26  &   0.07 &   0.08 & 62 &  49.52 (  0.12) &   1.65 (  0.28) &   0.10 (  0.94) &   0.14  &   0.06 &   0.05 \\ 
 6 &   6.03 (  0.09) &   1.61 (  0.21) &   0.10 (  0.95) &   0.21  &   0.07 &   0.07 & 64 &  10.83 (  0.02) &   3.14 (  0.05) &   0.42 (  0.19) &   0.93  &   0.04 &   0.05 \\ 
 8 &  32.56 (  0.01) &   3.36 (  0.01) &   0.77 (  0.06) &   3.13  &   0.07 &   0.21 & 65 &  85.12 (  0.01) &   3.04 (  0.03) &   1.80 (  0.14) &   2.29  &   0.08 &   0.12 \\ 
 12 & 111.03 (  0.03) &   2.92 (  0.08) &   0.35 (  0.31) &   0.95  &   0.06 &   0.09 & 66 &  66.04 (  0.02) &   2.53 (  0.05) &   1.26 (  0.24) &   0.78  &   0.04 &   0.04 \\ 
 13 &  21.88 (  0.01) &   3.53 (  0.01) &   3.41 (  0.02) &   3.01  &   0.06 &   0.24 & 67 &  32.83 (  0.02) &   4.13 (  0.04) &   0.30 (  0.12) &   1.64  &   0.06 &   0.07 \\ 
 14 &  33.23 (  0.01) &   1.88 (  0.02) &   0.90 (  0.13) &   0.69  &   0.02 &   0.02 & 68 &  55.11 (  0.01) &   2.80 (  0.01) &   0.10 (  0.01) &   1.08  &   0.06 &   0.06 \\ 
 15 &  59.96 (  0.01) &   2.48 (  0.01) &   0.10 (  0.01) &   2.44  &   0.06 &   0.08 & 70 &   7.05 (  0.07) &   2.08 (  0.18) &   0.10 (  0.51) &   0.32  &   0.07 &   0.07 \\ 
 16 &  59.76 (  0.01) &   3.09 (  0.01) &   0.11 (  0.01) &   4.00  &   0.06 &   0.15 & 71 &  14.05 (  0.04) &   2.29 (  0.13) &   2.44 (  0.78) &   0.45  &   0.06 &   0.08 \\ 
 17 &  45.12 (  0.01) &   1.86 (  0.01) &   1.11 (  0.07) &   3.73  &   0.06 &   0.13 & 72 &   3.11 (  0.01) &   3.93 (  0.02) &   0.10 (  0.04) &   1.94  &   0.06 &   0.08 \\ 
 18 & 120.93 (  0.01) &   2.87 (  0.02) &   0.10 (  0.02) &   1.91  &   0.06 &   0.09 & 73 &  69.89 (  0.03) &   3.09 (  0.04) &   0.73 (  0.01) &   1.21  &   0.10 &   0.10 \\ 
 19 &  43.73 (  0.01) &   2.81 (  0.01) &   0.22 (  0.01) &   5.33  &   0.06 &   0.12 & 74 &   5.05 (  0.01) &   2.93 (  0.03) &   0.10 (  0.05) &   2.04  &   0.09 &   0.09 \\ 
 20 &  34.85 (  0.06) &   1.57 (  0.16) &   0.10 (  0.17) &   0.28  &   0.06 &   0.07 & 75 &  22.95 (  0.01) &   1.20 (  0.02) &   2.83 (  0.20) &   2.47  &   0.06 &   0.10 \\ 
 21 &  34.14 (  0.03) &   1.10 (  0.11) &   1.03 (  1.18) &   0.65  &   0.08 &   0.08 & 79 &  22.52 (  0.01) &   1.80 (  0.01) &   0.91 (  0.03) &   6.78  &   0.04 &   0.18 \\ 
 22 &  84.56 (  0.01) &   2.45 (  0.04) &   1.54 (  0.20) &   2.91  &   0.12 &   0.15 & 80 &  29.42 (  0.01) &   1.74 (  0.03) &   0.55 (  0.22) &   1.50  &   0.06 &   0.07 \\ 
 23 &  84.25 (  0.01) &   2.14 (  0.03) &   0.68 (  0.16) &   2.84  &   0.08 &   0.13 & 82 &  20.42 (  0.02) &   1.55 (  0.06) &   4.14 (  0.76) &   0.61  &   0.04 &   0.06 \\ 
 25 &  77.96 (  0.01) &   2.59 (  0.01) &   0.10 (  0.01) &   4.71  &   0.09 &   0.21 & 85 &  11.21 (  0.03) &   1.74 (  0.12) &   0.34 (  0.60) &   0.49  &   0.05 &   0.06 \\ 
 27 &  76.74 (  0.01) &   2.78 (  0.03) &   0.26 (  0.10) &   3.99  &   0.09 &   0.21 & 87 &   5.55 (  0.01) &   0.82 (  0.02) &   1.84 (  0.52) &   1.11  &   0.05 &   0.06 \\ 
 28 &  84.29 (  0.01) &   2.26 (  0.03) &   1.88 (  0.17) &   2.67  &   0.08 &   0.18 & 88 &   5.85 (  0.01) &   0.64 (  0.02) &   2.36 (  0.84) &   0.82  &   0.05 &   0.05 \\ 
 29 &  58.89 (  0.02) &   2.57 (  0.04) &   1.96 (  0.25) &   1.58  &   0.08 &   0.14 & 90 &  -3.73 (  0.01) &   2.02 (  0.01) &   0.10 (  0.01) &   5.06  &   0.05 &   0.12 \\ 
 30 &  95.51 (  0.01) &   3.14 (  0.03) &   0.80 (  0.09) &   2.67  &   0.07 &   0.15 & 91 &  -2.04 (  0.01) &   0.95 (  0.03) &   2.20 (  0.74) &   0.62  &   0.05 &   0.05 \\ 
 33 & 109.70 (  0.01) &   2.70 (  0.01) &   0.10 (  0.01) &   2.83  &   0.07 &   0.10 & 93 &  -1.65 (  0.01) &   1.40 (  0.03) &   0.72 (  0.24) &   1.52  &   0.06 &   0.05 \\ 
 34 &  22.59 (  0.03) &   2.75 (  0.09) &   0.59 (  0.39) &   0.95  &   0.08 &   0.08 & 94 &   6.07 (  0.01) &   1.98 (  0.01) &   0.36 (  0.01) &   5.28  &   0.05 &   0.14 \\ 
 35 &  25.52 (  0.09) &   3.46 (  0.22) &   0.80 (  0.74) &   0.42  &   0.09 &   0.07 & 95 &   8.40 (  0.01) &   1.35 (  0.04) &   0.10 (  0.07) &   0.82  &   0.05 &   0.06 \\ 
 36 &  15.84 (  0.05) &   1.83 (  0.16) &   2.37 (  1.10) &   0.48  &   0.09 &   0.08 & 96 &  -2.99 (  0.02) &   1.84 (  0.06) &   0.59 (  0.35) &   1.22  &   0.07 &   0.08 \\ 
 37 & 105.17 (  0.02) &   1.94 (  0.07) &   1.94 (  0.50) &   1.14  &   0.09 &   0.09 & 97 &  -2.14 (  0.04) &   2.83 (  0.08) &   0.10 (  0.09) &   0.80  &   0.08 &   0.08 \\ 
 39 &  96.43 (  0.01) &   2.33 (  0.03) &   0.70 (  0.12) &   1.72  &   0.04 &   0.09 & 100 &  11.38 (  0.01) &   2.36 (  0.02) &   1.58 (  0.12) &   2.34  &   0.06 &   0.11 \\ 
 38 & 111.55 (  0.02) &   1.40 (  0.07) &   6.30 (  1.12) &   0.92  &   0.09 &   0.13 & 99 &  10.91 (  0.01) &   1.43 (  0.02) &   1.87 (  0.14) &   2.62  &   0.06 &   0.08 \\ 
 42 &  97.61 (  0.01) &   2.32 (  0.03) &   0.10 (  0.01) &   1.77  &   0.07 &   0.09 & 101 & -18.44 (  0.01) &   1.56 (  0.04) &   0.12 (  0.01) &   1.44  &   0.08 &   0.08 \\ 
 43 &  86.86 (  0.11) &   2.07 (  0.34) &   0.10 (  0.20) &   0.21  &   0.07 &   0.08 & 106 & -47.96 (  0.01) &   1.92 (  0.05) &   0.22 (  0.26) &   1.45  &   0.07 &   0.09 \\ 
 44 & 102.60 (  0.01) &   1.72 (  0.03) &   0.10 (  0.01) &   2.16  &   0.07 &   0.12 & 107 & -45.70 (  0.03) &   1.57 (  0.08) &   1.04 (  0.59) &   0.79  &   0.07 &   0.08 \\ 
 48 &  96.51 (  0.02) &   3.00 (  0.07) &   0.14 (  0.29) &   0.85  &   0.06 &   0.07 & 108 & -53.08 (  0.01) &   2.61 (  0.03) &   0.38 (  0.10) &   3.23  &   0.08 &   0.09 \\ 
 50 &  38.96 (  0.01) &   2.03 (  0.01) &   0.10 (  0.01) &   2.24  &   0.06 &   0.08 & 109 & -44.41 (  0.01) &   2.40 (  0.04) &   0.45 (  0.17) &   1.93  &   0.08 &   0.09 \\ 
 51 &  83.18 (  0.01) &   1.68 (  0.04) &   1.54 (  0.32) &   1.41  &   0.07 &   0.09 & 110 & -54.45 (  0.20) &   3.26 (  0.47) &   0.10 (  0.22) &   0.10  &   0.05 &   0.05 \\ 
 54 &  95.91 (  0.04) &   2.56 (  0.10) &   0.10 (  0.08) &   0.78  &   0.09 &   0.09 & 111 & -17.54 (  0.03) &   1.22 (  0.08) &   0.84 (  0.82) &   0.61  &   0.07 &   0.07 \\ 
 55 &  49.65 (  0.01) &   3.17 (  0.02) &   0.10 (  0.02) &   2.26  &   0.07 &   0.08 & 112 & -18.83 (  0.09) &   0.92 (  0.25) &   0.10 ( 19.42) &   0.16  &   0.07 &   0.08 \\ 
 59 &  77.18 (  0.02) &   2.54 (  0.04) &   0.11 (  0.01) &   1.56 &   0.07 &   0.10 \\ 
 \hline
\end{tabular}}
\end{center}
\end{sidewaystable}


\begin{sidewaystable}
\begin{center}
\caption{Results from Gaussian fits to \hthirteencoplus\ spectra.
  For each source the table gives (in order) the velocity of the emission, the
  velocity width of the line, the integrated intensity $\int T_A^*\,dv$, the
  line peak temperature in units of $T_A^*$ from the fit, the rms in an
  emission-free region of the spectrum and the rms residual in the region of
  the line after fitting the Gaussian. The values in parenthesis are the one
  standard deviation uncertainties in the associated quantities.
}\label{tab:h13cofits} {\small
\begin{tabular}{ccccccc|ccccccc}
\hline \hline
Source & V & dv & Int. I. & T & Baseline & Line & Source & V & dv & Int. I & T & Baseline & Line \\
               &  &    &     &      & rms  & rms & &  &    &     &      & rms  & rms \\
               & (km/s) & (km/s) & (K km/s) & (K) & (K) & (K) & & (km/s) & (km/s) & (K km/s) & (K) & (K) & (K) \\
\hline


2 & -16.09 (  0.03) &   2.46 (  0.08) &   2.74 (  0.08) &   1.05 &   0.11 &   0.12 & 71 &  14.06 (  0.11) &   2.44 (  0.27) &   0.37 (  0.03) &   0.14 &   0.05 &   0.03 \\ 
 3 &   0.13 (  0.08) &   1.38 (  0.19) &   0.55 (  0.06) &   0.37 &   0.12 &   0.12 & 72 &   2.75 (  0.09) &   5.16 (  0.21) &   2.53 (  0.09) &   0.46 &   0.08 &   0.07 \\ 
 6 &   5.67 (  0.05) &   1.60 (  0.11) &   0.97 (  0.06) &   0.57 &   0.11 &   0.10 & 75 &  22.75 (  0.04) &   1.75 (  0.11) &   1.47 (  0.08) &   0.79 &   0.12 &   0.11 \\ 
 8 &  32.96 (  0.03) &   4.05 (  0.07) &   4.91 (  0.07) &   1.14 &   0.08 &   0.15 & 79 &  22.23 (  0.03) &   2.24 (  0.09) &   2.79 (  0.09) &   1.17 &   0.13 &   0.11 \\ 
 13 &  21.79 (  0.09) &   3.73 (  0.20) &   1.58 (  0.07) &   0.40 &   0.08 &   0.08 & 80 &  29.16 (  0.09) &   1.96 (  0.19) &   0.83 (  0.08) &   0.40 &   0.13 &   0.11 \\ 
 14 &  33.27 (  0.02) &   2.13 (  0.04) &   3.70 (  0.05) &   1.63 &   0.08 &   0.09 & 82 &  19.99 (  0.11) &   1.72 (  0.24) &   0.46 (  0.06) &   0.25 &   0.05 &   0.04 \\ 
 16 &  59.47 (  0.05) &   3.73 (  0.10) &   3.34 (  0.08) &   0.84 &   0.09 &   0.12 & 85 &  11.08 (  0.10) &   2.11 (  0.23) &   0.62 (  0.06) &   0.28 &   0.09 &   0.07 \\ 
 17 &  45.15 (  0.04) &   2.46 (  0.10) &   1.81 (  0.06) &   0.69 &   0.08 &   0.09 & 88 &   5.84 (  0.03) &   0.70 (  0.06) &   0.29 (  0.02) &   0.39 &   0.06 &   0.06 \\ 
 18 & 120.71 (  0.10) &   3.03 (  0.24) &   2.02 (  0.14) &   0.63 &   0.18 &   0.19 & 87 &   5.54 (  0.03) &   0.90 (  0.07) &   0.32 (  0.02) &   0.33 &   0.06 &   0.05 \\ 
 19 &  43.82 (  0.02) &   3.16 (  0.06) &   9.33 (  0.14) &   2.77 &   0.16 &   0.20 & 90 &  -3.46 (  0.02) &   3.09 (  0.06) &   4.80 (  0.07) &   1.46 &   0.09 &   0.09 \\ 
 21 &  34.38 (  0.07) &   0.76 (  0.29) &   0.20 (  0.04) &   0.25 &   0.07 &   0.05 & 91 &  -2.03 (  0.03) &   1.14 (  0.08) &   0.74 (  0.04) &   0.61 &   0.09 &   0.11 \\ 
 23 &  84.03 (  0.09) &   2.73 (  0.19) &   0.71 (  0.04) &   0.24 &   0.06 &   0.06 & 93 &  -1.74 (  0.04) &   1.39 (  0.08) &   1.17 (  0.06) &   0.79 &   0.12 &   0.10 \\ 
 25 &  78.01 (  0.11) &   2.39 (  0.30) &   0.84 (  0.08) &   0.33 &   0.11 &   0.11 & 94 &   6.05 (  0.05) &   2.11 (  0.11) &   1.36 (  0.06) &   0.60 &   0.09 &   0.10 \\ 
 28 &  83.73 (  0.12) &   2.14 (  0.31) &   0.67 (  0.08) &   0.30 &   0.12 &   0.13 & 95 &   8.32 (  0.06) &   1.68 (  0.17) &   0.69 (  0.06) &   0.39 &   0.09 &   0.08 \\ 
 30 &  95.52 (  0.11) &   2.33 (  0.27) &   1.00 (  0.09) &   0.40 &   0.13 &   0.11 & 99 &  10.94 (  0.03) &   1.66 (  0.07) &   1.94 (  0.07) &   1.09 &   0.12 &   0.11 \\ 
 35 &  25.86 (  0.17) &   4.88 (  0.42) &   1.59 (  0.12) &   0.31 &   0.12 &   0.11 & 100 &  11.40 (  0.05) &   2.66 (  0.10) &   2.40 (  0.09) &   0.85 &   0.12 &   0.10 \\ 
 48 &  96.38 (  0.10) &   1.73 (  0.26) &   0.74 (  0.09) &   0.40 &   0.14 &   0.13 & 101 & -18.38 (  0.05) &   1.39 (  0.12) &   0.91 (  0.07) &   0.62 &   0.12 &   0.11 \\ 
 65 &  85.55 (  0.15) &   3.72 (  0.36) &   1.99 (  0.16) &   0.50 &   0.18 &   0.19 & 108 & -53.52 (  0.11) &   3.00 (  0.27) &   1.18 (  0.09) &   0.37 &   0.11 &   0.09 \\ 
 66 &  65.69 (  0.06) &   3.21 (  0.15) &   1.60 (  0.06) &   0.47 &   0.07 &   0.08 & 109 & -44.13 (  0.07) &   2.95 (  0.17) &   2.05 (  0.10) &   0.65 &   0.13 &   0.12 \\ 
 67 &  33.07 (  0.11) &   3.90 (  0.27) &   1.46 (  0.09) &   0.35 &   0.10 &   0.09 & 111 & -17.68 (  0.06) &   1.58 (  0.20) &   0.85 (  0.08) &   0.50 &   0.12 &   0.12 \\ 
 68 &  54.90 (  0.09) &   2.40 (  0.17) &   0.72 (  0.05) &   0.28 &   0.07 &   0.09 & 112 & -18.76 (  0.06) &   1.26 (  0.16) &   0.62 (  0.06) &   0.46 &   0.12 &   0.13 \\ 
 71 &  14.06 (  0.11) &   2.44 (  0.27) &   0.37 (  0.03) &   0.14 &   0.05 &   0.03 \\ 
 \hline
\end{tabular}}
\end{center}
\end{sidewaystable}


\section{Line Asymmetries}
\label{sec:asymm}

To quantify the blue asymmetry of a line we use the \citet{mar} asymmetry
parameter, $\delta \textrm{v}$, which is defined as the difference between the
line peak velocity of an optically thick tracer, $\textrm{v}(\textrm{thick})$,
and an optically thin tracer, $\textrm{v}(\textrm{thin})$, measuring the
systemic velocity, in units of the optically thin line width,
$\textrm{dv}(\textrm{thin})$,
\begin{eqnarray}
\delta \textrm{v} &=& \frac{\textrm{v}(\textrm{thick})-\textrm{v}(\textrm{thin})}{\textrm{dv}(\textrm{thin})}
\end{eqnarray}
where we use \ntwohplus\ as the optically thin tracer. 

\begin{figure*}
  \centering
  \caption{Asymmetry for the four transitions for the sample of objects with
    known distances of $<$6 kpc. The dashed lines mark the $|\delta
    \textrm{v}|<0.25$ region which defines the spectra with no asymmetry.}
  \label{fig:asymm}
  \includegraphics[angle=-90, width=14cm]{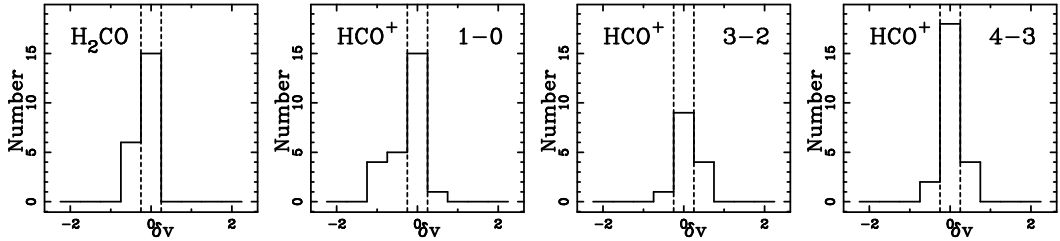}
\end{figure*}

Table~\ref{tab:asymm} and Figure~\ref{fig:asymm} summarize the asymmetry
parameter measurements for the \hcoplus\ \j{1}{0}, \j{3}{2}, \j{4}{3} and
\htwoco\ spectra, measured with respect to the \ntwohplus\ velocity and line
width. Negative values correspond to blueshifted emission while positive
values are redshifted. For comparison, Table~\ref{tab:asymm} also includes the
\ntwohplus\ and \hthirteencoplus\ line velocities as measured by hyperfine
fitting and Gaussian fitting respectively, together with the velocity of the
\hcoplus\ \j{4}{3} transition also derived from a Gaussian fit to the line.
Comparing the results from the Gaussian and parabolic fits shows that for the
lines which are most Gaussian in shape the two methods produce very good
agreement in the fitted velocity. Examination of those eight sources where the
results from the two methods differ by more than 0.3\kms\ shows that the line
profiles are clearly double peaked or skewed. The rms difference in velocities
over all the sources is 0.26\kms\ with a mean difference of 0.03\kms, while if
only those 65 sources with a velocity difference of less than 0.5\kms\ are
considered the rms velocity difference drops to 0.14\kms\ with a mean velocity
difference of -0.02\kms. This good correspondence between the Gaussian and
peak fitted velocities, especially for sources with nearly Gaussian lines
confirms that the peak fitting method produces reliable estimates of the line
peak velocity.

\begin{sidewaystable}
\caption{Line velocities and asymmetries. Columns 2, 3 and 4 give the
  line velocities determined from the hyperfine fit to the \ntwohplus\ and
  Gaussian fits to the \hthirteencoplus\ and \hcoplus\ \j{4}{3} respectively.
  The next four columns give the velocities derived from parabolic fits to the
  lines and the final four columns the line asymmetry parameter derived from
  these velocities.}\label{tab:asymm} \centering
\begin{tabular}{lcccccccccccc}
\hline\hline
Source & v(\ntwohplus) &  v(\hthirteencoplus) & v(\hcoplus\ 4-3)& \multicolumn{4}{c}{Velocity} & &\multicolumn{4}{c}{$\delta \textrm{v}$}\\ \cline{5-8} \cline {10-13} 
 & \ldots & \ldots &  \ldots &  \hcoplus(1-0) & \hcoplus(3-2) & \hcoplus(4-3)
& \htwoco\ & &\hcoplus(1-0) & \hcoplus(3-2) & \hcoplus(4-3) & \htwoco\\ \hline
2  & -16.10 &   -16.09 &   -14.98 &   -15.49 & \ldots &    -14.85 & \ldots & &     0.23 &  \ldots &      0.46 & \ldots \\ 
3  &   0.22 &     0.13 &     0.27 &     0.12 & \ldots &      0.32 & \ldots &&     -0.07 &  \ldots &      0.07 & \ldots \\ 
5  &   0.72 & \ldots & \ldots &      0.19 & \ldots & \ldots &  \ldots &   &  -0.53 & \ldots & \ldots & \ldots \\ 
6  &   6.03 &     5.67 &     5.83 &     6.19 & \ldots &      5.83 & \ldots &    &  0.10 &  \ldots &     -0.12 & \ldots \\ 
8  &  32.56 &    32.96 &    31.57 &    35.29 &    31.26 &    31.59 &    34.84 & &    0.81&     -0.39 &    -0.29 &     0.68 \\
12  & 111.03 & \ldots &    110.86 &   111.72 & \ldots &    110.38 &   111.60 &  &   0.23 &  \ldots &     -0.22 &     0.20 \\
13  &  21.88 &    21.79 &    21.96 &    24.80 & \ldots &     23.42 &    22.99 & &    0.83 & \ldots &      0.44 &     0.31 \\
14  &  33.23 &    33.27 &    33.24 &    33.26 & \ldots &     33.21 &    33.25 & &    0.02 & \ldots &     -0.01 &     0.01 \\
15  &  59.96 & \ldots &     59.71 &    60.80 & \ldots &     59.79 &    60.49 &  &   0.34 &  \ldots &     -0.07 &     0.22 \\
16  &  59.76 &    59.47 &    58.77 &    58.41 & \ldots &     58.88 &    58.66 & &   -0.44 & \ldots &     -0.29 &    -0.36 \\
17  &  45.12 &    45.15 &    45.41 &    44.54 & \ldots &     45.19 &    44.95 & &   -0.31 & \ldots &      0.04 &    -0.09 \\
18  & 120.93 &   120.71 &   120.66 &   120.49 & \ldots &    120.66 &   120.54 & &   -0.15 & \ldots &     -0.10 &    -0.14 \\
19  &  43.73 &    43.82 &    44.35 &    44.01 & \ldots &     44.87 &    43.79 & &    0.10 & \ldots &      0.40 &     0.02 \\
21  &  34.14 &    34.38 &    34.04 &    32.78 & \ldots &     34.17 &    33.96 & &   -1.23 & \ldots &      0.03 &    -0.16 \\
20  &  34.85 & \ldots &     34.73 &    32.95 & \ldots &     34.78 &    33.95 &  &  -1.20 &  \ldots &     -0.04 &    -0.57 \\
22  &  84.56 & \ldots &     83.75 &    84.76 &    83.74 &    83.63 &    84.17 & &    0.08 &    -0.34 &    -0.38 &    -0.16 \\
23  &  84.25 &    84.03 &    84.17 &    84.67 & \ldots &     84.19 &    84.20 & &    0.20 & \ldots &     -0.03 &    -0.02 \\
25  &  77.96 & \ldots &     77.98 &    77.68 & \ldots &     77.87 &    77.76 &  &  -0.11 &  \ldots &     -0.03 &    -0.08 \\
27  &  76.74 & \ldots &     76.78 &    77.45 & \ldots &     77.48 &    76.98 &  &   0.26 &  \ldots &      0.27 &     0.09 \\
28  &  84.29 & \ldots &     84.13 &    82.79 & \ldots &     84.89 &    83.65 &  &  -0.66 &  \ldots &      0.27 &    -0.29 \\
29  &  58.89 & \ldots &     59.27 &    57.56 & \ldots &     58.87 &    58.82 &  &  -0.52 &  \ldots &     -0.01 &    -0.03 \\
30  &  95.51 &    95.52 &    96.12 &    97.30 & \ldots &     96.28 &    95.73 & &    0.57 & \ldots &      0.25 &     0.07 \\
33  & 109.70 & \ldots &    110.35 &   110.48 & \ldots &    110.46 &   110.49 &  &   0.29 &  \ldots &      0.28 &     0.29 \\
34  &  22.59 & \ldots &     22.95 &    20.97 & \ldots &     22.93 &    22.31 &  &  -0.59 &  \ldots &      0.12 &    -0.10 \\
35  &  25.52 & \ldots &     25.98 &    27.56 & \ldots &     25.97 &    25.25 &  &   0.59 &  \ldots &      0.13 &    -0.08 \\
36  &  15.84 & \ldots &     15.81 &    16.10 & \ldots &     15.85 &    15.10 &   &  0.14 &  \ldots &      0.00 &    -0.41 \\
37  & 105.17 & \ldots &    105.39 &   105.60 & \ldots &    105.36 &   105.14 &   &  0.22 &  \ldots &      0.10 &    -0.02 \\
39  &  96.43 & \ldots &     95.59 &    95.50 & \ldots &     95.68 &    95.53 &   & -0.40 &  \ldots &     -0.32 &    -0.39 \\
38  & 111.55 & \ldots &    111.47 &   111.53 & \ldots &    111.45 &   111.30 &  &  -0.02 &  \ldots &     -0.07 &    -0.18 \\
42  &  97.61 & \ldots &     97.97 &    97.56 & \ldots &     97.99 &    97.63 &  &  -0.02 &  \ldots &      0.16 &     0.01 \\
44  & 102.60 & \ldots &    102.65 &   102.65 & \ldots &    102.66 &   102.34 &    & 0.03 &  \ldots &      0.04 &    -0.15 \\
48  &  96.51 & \ldots &     97.07 & \ldots & \ldots &  \ldots & \ldots & & \ldots & \ldots&   \ldots & \ldots \\ 
50  &  38.96 & \ldots &     38.98 &    38.65 &    38.75 &    38.94 &    38.69 &   & -0.15&     -0.10 &    -0.01 &    -0.13 \\
51  &  83.18 & \ldots &     83.17 &    82.10 & \ldots &     83.15 &    82.68 &   & -0.64 &  \ldots &     -0.01 &    -0.30 \\
55  &  49.65 & \ldots &     49.51 &    47.86 & \ldots &     49.56 &    48.56 &   & -0.57 &  \ldots &     -0.03 &    -0.34 \\
59  &  77.18 & \ldots &     77.22 &    74.92 &    76.08 &    76.84 &    76.15 &  &  -0.89&     -0.43 &    -0.13 &    -0.41 \\
\hline
\end{tabular}
\end{sidewaystable}
\begin{sidewaystable*}
\begin{flushleft}
{Table 7 continued.\\}
\end{flushleft}
\centering
\begin{tabular}{lcccccccccccc}
\hline\hline
Source & v(\ntwohplus) &  v(\hthirteencoplus) & v(\hcoplus\ 4-3)& \multicolumn{4}{c}{Velocity} && \multicolumn{4}{c}{$\delta \textrm{v}$}\\ \cline{5-8} \cline {10-13} 
 & \ldots & \ldots &  \ldots &  \hcoplus(1-0) & \hcoplus(3-2) & \hcoplus(4-3)
& \htwoco\ && \hcoplus(1-0) & \hcoplus(3-2) & \hcoplus(4-3) & \htwoco\\ \hline
62  &  49.52 & \ldots &     49.70 &    50.20 & \ldots &     49.70 &    49.91 &   &  0.41 & \ldots &      0.11 &     0.24 \\
64  &  10.83 & \ldots &     10.15 &     9.04 &     9.84 &    10.11 &     9.70 &   & -0.57 &    -0.32 &    -0.23 &    -0.36 \\
65  &  85.12 & \ldots &     87.45 &    88.00 &    87.58 &    87.40 &    87.00 & &    0.95 &     0.81 &     0.75 &     0.62 \\
66  &  66.04 &    65.69 &    65.51 &    65.97 &    64.38 &    65.11 &    64.93 &&    -0.03 &    -0.65 &    -0.37 &    -0.44 \\
67  &  32.83 &    33.07 &    32.94 &    32.38 &    32.41 &    32.68 &    32.71 &&    -0.11 &    -0.10 &    -0.04 &    -0.03 \\
68  &  55.11 &    54.90 &    54.94 &    54.81 &    55.15 &    54.94 &    54.91 &&    -0.11 &     0.01 &    -0.06 &    -0.07 \\
71  &  14.05 &    14.06 &    14.38 &    14.58 & \ldots &     14.31 &    14.47 & &    0.23 & \ldots &      0.12 &     0.18 \\
70  &   7.05 & \ldots &      6.45 &     6.99 &     7.17 &     6.40 &     6.94 & &   -0.03 &     0.06 &    -0.32 &    -0.05 \\
72  &   3.11 &     2.75 &     3.43 &     3.45 & \ldots &      3.48 &     3.18 & &    0.09 & \ldots &      0.09 &     0.02 \\
73  &  69.89 & \ldots &     69.14 &    68.65 &    68.59 &    69.04 &    68.72 & &   -0.40 &    -0.42 &    -0.28 &    -0.38 \\
74  &   5.05 & \ldots & \ldots &      5.10 & \ldots & \ldots &      5.14 & &    0.02 & \ldots & \ldots &      0.03 \\
75  &  22.95 &    22.75 &    22.89 &    22.74 & \ldots &     22.88 &    22.90 &  &  -0.18 & \ldots &     -0.06 &    -0.04 \\
79  &  22.52 &    22.23 &    21.85 &    21.38 & \ldots &     21.72 &    22.06 &  &  -0.64 & \ldots &     -0.44 &    -0.26 \\
80  &  29.42 &    29.16 &    29.76 &    29.43 & \ldots &     29.90 &    29.49 &  &   0.00 & \ldots &      0.27 &     0.04 \\
82  &  20.42 &    19.99 &    20.04 &    19.69 & \ldots &     19.74 &    19.77 &  &  -0.48 & \ldots &     -0.44 &    -0.42 \\
85  &  11.21 &    11.08 &    11.03 &    12.49 & \ldots &     11.11 &    12.13 &  &   0.73 & \ldots &     -0.06 &     0.53 \\
88  &   5.85 &     5.84 &     5.63 &     5.44 & \ldots &      5.61 &     5.65 &  &  -0.65 & \ldots &     -0.37 &    -0.32 \\
87  &   5.55 &     5.54 &     5.51 &     4.79 & \ldots &      5.50 &     5.52 &  &  -0.93 & \ldots &     -0.06 &    -0.04 \\
90  &  -3.73 &    -3.46 &    -3.18 &    -3.27 & \ldots &     -2.94 &    -3.55 &  &   0.23 & \ldots &      0.39 &     0.09 \\
91  &  -2.04 &    -2.03 &    -1.79 &    -2.20 & \ldots &     -1.79 &    -1.83 &  &  -0.17 & \ldots &      0.26 &     0.21 \\
93  &  -1.65 &    -1.74 &    -1.45 &    -1.60 &    -1.46 &    -1.54 &    -1.63 & &    0.03 &     0.13 &     0.07 &     0.01 \\
94  &   6.07 &     6.05 &     6.28 &     6.24 &     6.52 &     6.23 &     6.29 & &    0.09 &     0.23 &     0.08 &     0.11 \\
95  &   8.40 &     8.32 &     8.55 &     8.59 &     8.68 &     8.58 &     8.49 & &    0.14 &     0.21 &     0.14 &     0.07 \\
96  &  -2.99 & \ldots &     -2.66 &    -3.75 &    -2.77 &    -2.68 &    -3.55 &  &  -0.41 &     0.12 &     0.17 &    -0.30 \\
97  &  -2.14 & \ldots &     -1.66 &    -2.67 &    -1.31 &    -1.62 &    -1.69 &  &  -0.19 &     0.29 &     0.18 &     0.16 \\
99  &  10.91 &    10.94 &    10.92 &    11.12 &    10.73 &    10.96 &    11.16 & &    0.15 &    -0.12 &     0.03 &     0.17 \\
100  &  11.38 &    11.40 &    11.34 &    11.41 &    11.22 &    11.32 &    11.48 & &    0.01 &    -0.07 &    -0.03 &     0.04 \\
101  & -18.44 &   -18.38 &   -18.20 &   -17.40 &   -17.50 &   -18.18 &   -18.84 & &    0.66 &     0.60 &     0.16 &    -0.26 \\
106  & -47.96 & \ldots &    -47.99 &   -48.13 &   -47.71 &   -47.93 &   -47.74 &  &  -0.09 &     0.13 &     0.01 &     0.11 \\
107  & -45.70 & \ldots &    -45.54 &   -45.55 &   -45.60 &   -45.53 &   -45.74 &  &   0.10 &     0.06 &     0.11 &    -0.02 \\
108  & -53.08 &   -53.52 &   -52.83 &   -53.27 &   -52.48 &   -52.58 &   -53.05 & &   -0.07 &     0.23 &     0.19 &     0.01 \\
109  & -44.41 &   -44.13 &   -43.32 &   -45.07 &   -43.08 &   -43.10 &   -44.67 & &   -0.27 &     0.56 &     0.55 &    -0.11 \\
111  & -17.54 &   -17.68 &   -17.65 &   -18.65 &   -17.16 &   -17.70 &   -17.86 &  &  -0.91 &     0.31 &    -0.13 &    -0.27 \\
\hline
\end{tabular}
\end{sidewaystable*}


Most mechanisms producing asymmetric line profiles towards sources, e.g.
rotation, should produce approximately equal numbers of red and blue
asymmetric profiles, there being no mechanism to favour one colour of
asymmetry over the other.  On the other hand infall should preferentially
produce blue asymmetric profiles (e.g. Anglada et al. 1987; Zhou 1992; Walker,
Narayanan \& Boss 1994).  To quantify whether line profiles of a particular
colour dominate in a sample, \citet{mar} defined a quantity $\tens{E}$, the
blue excess,
\begin{eqnarray}
  \label{eq:blueex}
  \tens{E}  &=& \frac{N_{\rm blue} - N_{\rm red}}{N_{\rm total}}
\end{eqnarray}
where $N_{\rm blue}$ and $N_{\rm red}$ are the number of sources which
show blue or red asymmetric lines, respectively, and $N_{\rm total}$
is the total number of sources observed.  The presence of large blue
excesses towards samples of low mass protostars has provided
considerable support to the case for the detection of infall around
such sources \citep{mar}.

To apply the blue excess statistic requires the definition of a range of
$\delta \textrm{v}$ to identify asymmetric sources. Mardonnes et al. adopted a
criterion of $|\delta \textrm{v}|>0.25$ to indicate that a line profile was
asymmetric and for consistency with their analysis we adopt the same
criterion. It is important to note however that since the \ntwohplus\ lines
towards this sample of sources is significantly broader than towards low mass
sources, the choice of this same bound on $\delta \textrm{v}$ actually
corresponds to a larger absolute velocity shift for these high mass objects
than for the lower mass sources.  Combining the estimated uncertainties on the
measured line velocities and widths we estimate a typical uncertainty in our
measured values of $\delta \textrm{v}$ of 0.02 to 0.05.

The results of measuring the blue excess for the four transitions which are
assumed to be optically thick are listed in Table~\ref{tab:blueex}.  This
table shows that for the whole sample of sources observed in \hcoplus\ 
\j{3}{2}\ and \j{4}{3}\ have an equal number of blue and red asymmetric lines.
However in both the \j{1}{0}\ transiton of \hcoplus\ and \htwoco, there are
considerably more blue asymmetric lines than red asymmetric ones, with blue
excesses of 15\% and 19\% for the \hcoplus\ and \htwoco\ respectively.  In
terms of the actual number of sources with significant asymmetries for
\hcoplus\ \j{1}{0}\ there are nearly twice as many blue sources as red sources
while for \htwoco\ there are over three times as many blue lines as red ones.

If is of course possible that for any set of observations an excess of blue
sources could arise by chance even if the underlying distribution of sources
has equal number of red and blue asymmetric line profiles.  So in order to
assess the statistical significance of the measured blue excesses we have used
an exact probability binomial test \citep{con}.  Given that we have a number
of sources which have a significant red asymmetry ($\delta \textrm{v}> 0.25$) or blue
asymmetry ($\delta \textrm{v}<-0.25$), if the sources are randomly distributed between
red and blue asymmetric, we can calculate the probability that we would
observe by chance a blue excess as large as actually measured.  We do this by
calculating the number of ways it is possible to distribute a number of
objects equal to the total of blue plus red asymmetric profiles into two bins,
blue and red, and ask how many of these combinations have a number of sources
in the blue bin equal to or larger than observed.  The ratio of this to the
total number of possible combinations is then the probabilty that the observed
excess arises by chance. This probability is listed in Table~\ref{tab:blueex}
as the quantity ${\rm P}$.  A small value of ${\rm P}$ indicates that it is
unlikely that a blue excess as large as observed could arise by chance from an
intrinsically uniform distribution.


\begin{table*}[h]
\caption{Blue excess statistics. The table shows for each transition 
the number of sources with blue asymmetric lines, red asymmetric lines 
and the total number of sources, the blue
  excess, $\tens{E}$, the probability, P, of such an excess and the mean
  asymmetry and the standard error on the mean.}\label{tab:blueex}
\centering
\begin{tabular}{lcccrcr}
\hline\hline
Transition  &  $N_{\rm blue}$ & $N_{\rm red}$ & $N_{\rm total}$ &
$\tens{E}$ & P&  Mean $\delta \textrm{v}\pm$sem\\ \hline
\hcoplus\ \j{1}{0} & 21 & 11 & 68 & 0.15 & 0.06 &$-0.08\pm0.05$  \\
\hcoplus\ \j{3}{2} & 6  & 5  & 24 & 0.04 & 0.50  & $0.03\pm0.07$  \\
\hcoplus\ \j{4}{3} & 10 & 11 & 66 & -0.02 & 0.67 & $0.03\pm0.03$ \\
\htwoco            & 17 & 5  & 64 & 0.19 & 0.008 & $-0.05\pm0.03$  \\
\multicolumn{7}{c}{Distance $<6$kpc subsample }\\
\hcoplus\ \j{1}{0} & 9 & 1 & 25 & 0.32 & 0.01 & $-0.22\pm0.09$\\
\hcoplus\ \j{3}{2} & 1  & 4 & 14 & $-$0.21 & 0.97 & $0.14\pm0.07$\\
\hcoplus\ \j{4}{3} & 2 & 4  & 24 & $-0.08$ & 0.89 & $0.07\pm0.04$ \\
\htwoco            & 6 & 0  & 21 & 0.29 & 0.02 & $-0.07\pm0.04$\\
\hline
\end{tabular}
\end{table*}


Table~\ref{tab:blueex} shows that for \hcoplus\ \j{1}{0} the excess of blue
asymmetric sources as compared to red, the binomial test indicates only a 6\%
probability of this asymmetry arising by chance.  For the \htwoco\ with a
total number of asymmetric sources is 22, 17 blue and 5 red, the probability
is just 0.8\% that this excess arises by chance. It appears that for these two
transitions, but particularly the \htwoco, the excess of blue asymmetric lines
is statistically significant.

One point to recall about the sources in this sample is that they span a range
of distances from 0.7 to 10.6 kpc \citep{tk1} and for some objects there is an
ambiguity between the source near and far kinematic distances leading to the
possibility that some sources are at distants as great as 13.6 kpc
\citep{tk1}. To limit the possibility of including sources at very large
distances where the telescope beam may contain emission from material not
intimately associated with the high mass protostars, we formed a subsample of
objects which have confirmed distances of less than 8kpc and have looked at
the blue excess for these objects alone.  The median distance of the sources
in this subsample is 3.3 kpc.

For this distance limited subsample the blue excess seen in \hcoplus\ 
\j{1}{0}\ and \htwoco\ become even more pronounced with blue excess of
28\% and 23\% respectively and probabilities of only 2\% and 3\% that
these arise by chance.  Restricting the sample even further to include
only sources with known distances of less than 6 kpc, giving a mean
distance of a source of 2.7kpc, the blue excesses become even more
significant (Table~\ref{tab:blueex}).  In the \hcoplus\ \j{1}{0} the
sample falls to 25 sources, 9 blue and 1 red, giving a blue excess of
0.31 and a mean asymmetry of $-0.22\pm0.09$ while for the \htwoco\ the
sample falls to 21 sources, 6 blue and none red, giving an excess of
0.32 with a mean asymmetry of $-0.07\pm0.04$. The associated
probabilities drop to 1\% and 2\% respectively.

In comparing the statistical significance of the blue excesses in this survey
with other surveys for infall it is important to note that other surveys have
not used this exact binomial test which we have used. For example, \citet{mar}
used the Student-t test to compare the distribution of $\delta \textrm{v}$
with a normal distrbution with zero mean.  This kind of test makes an
assumption about the nature of the underlaying distribution of asymmetries,
including those sources which do not show any asymmetry.  However there are
many reasons why a source may not show an infall asymmetry which do not fit
well with the hypothesis of smooth distribution of asymmetries assumed by the
Student-t test. For example a source might be at a slightly different
evolutionary stage, or might not quite have the excitation conditions
necessary to produce the asymmetry. We therefore prefer the binomial test
which, once stating a criterion which defines which sources have asymmetric
profiles, only considers those sources. However, for the sake of comparison
applying the Student t-test to just those sources which have a measured
asymmetry (which is not how, for example, \cite{mar} used the test) does
produce probabilities similar to but somewhat larger than those given by our
binomial test. For example, for the \hcoplus\ \j{1}{0}\ data on the whole
sample, the Student t-test gives a probability of the observed distribution of
blue and red sources arising by chance of 0.08, similar to the 0.06 from the
binomial test.  Applied to the whole sample of asymmetry parameters for the
same line in the $<6$kpc subsample, the Student t-test gives a probability of
0.025, compared to 0.01 for the binomial test applied to only the asymmetric
lines.


As mentioned in Section~\ref{sec:red}, there are a number of sources
for which the residuals after fitting the simple single velocity
component model for the \ntwohplus\ suggest the presence of additional
emission. However the presence of these sources in the sample do not
significantly distort the measured blue excess. To demonstrate this we
have excluded from the sample all the sources for which the rms of the
single component fit is greater than 1.6 times the rms in the line
free region of the spectra and recalculated the blue excess
statistics. For the remaining sources there are 15 blue and 7 red out
of 50 \hcoplus\ \j{1}{0}\ spectra giving a blue excess of 0.16 with an
associated probability of 7\%, while for the 47 remaining \htwoco\ 
spectra, there are 13 blue and 3 red giving a blue excess of 0.21 with
a probability of 1\%. Comparing these values with those in
Table~\ref{tab:blueex} shows that the relatively poorly fitted sources
are having negligible effect on the statistics. The same is true if
the 6kpc subsample is examined.

For both the 8kpc and 6kpc samples, both \hcoplus\ \j{4}{3}\ and \j{3}{2}\ 
show relatively large {\it negative} blue excesses, in other words, an excess
of red asymmetric lines. However in no case does the probability show this
excess to be less than a 11\% probability of being due to chance and so these
red excesses appear consistent with chance variations due to sampling.

\begin{table*}[bh]
\caption{Summary of sources with at least one significant line
  asymmetry. Entries marked R or B indicate
  red and blue asymmetric line profile respectively. A dash (--)
  indicates a line profile with no significant asymmetry and the
  absence of an entry indicates no observation. Sources marked  * are
  considered the best infall candidates. }\label{tab:linesum}
\begin{center}
\begin{tabular}{lcccc|lcccc}
\hline\hline
Source &  \multicolumn{3}{c}{\hcoplus} & \htwoco &
Source &  \multicolumn{3}{c}{\hcoplus} & \htwoco 
\\ \cline{2-4}\cline{7-9}  
 &   $1\rightarrow0$ & $3\rightarrow2$ & $4\rightarrow3$ & & 
 &   $1\rightarrow0$ & $3\rightarrow2$ & $4\rightarrow3$  \\ \hline
2 &  --  &      &  R   &       &   55 *  &  B   &      &  --  &  B    \\ 
5 * &  B   &      &      &       &   59 * &  B   &  B   &  --  &  B    \\ 
8 &  R   &  B   &  B   &  R    &   62 &  R   &      &  --  &  --   \\ 
13 &  R   &      &  R   &  R   &   64 * &  B   &  B   &  --  &  B    \\ 
15 &  R   &      &  --  &  --  &   65 &  R   &  R   &  R   &  R    \\ 
16 *  &  B   &      &  B   &  B   &   66 * &  --  &  B   &  B   &  B    \\ 
17 * &  B   &      &  --  &  --  &   70 * &  --  &  --  &  B   &  --   \\ 
19 &  --  &      &  R   &  --  &   73 * &  B   &  B   &  B   &  B    \\ 
21 * &  B   &      &  --  &  --  &   79 * &  B   &      &  B   &  B    \\ 
20 * &  B   &      &  --  &  B   &   80 &  --  &      &  R   &  --   \\ 
22 * &  --  &  B   &  B   &  --  &   82 * &  B   &      &  B   &  B    \\ 
27 &  R   &      &  R   &  --  &   85 &  R   &      &  --  &  R    \\ 
28 &  B   &      &  R   &  B   &   87 * &  B   &      &  --  &  --   \\ 
29 * &  B   &      &  --  &  --  &   88 * &  B   &      &  B   &  B    \\ 
30 &  R   &      &  --  &  --  &   90 &  --  &      &  R   &  --   \\ 
33 &  R   &      &  R   &  R   &   91 &  --  &      &  R   &  --   \\ 
34 * &  B   &      &  --  &  --  &   96 * &  B   &  --  &  --  &  B    \\ 
35 &  R   &      &  --  &  --  &   97 &  --  &  R   &  --  &  --   \\ 
36 * &  --  &      &  --  &  B   &   101 &  R   &  R   &  --  &  B    \\
39 * &  B   &      &  B   &  B   &   109 &  B   &  R   &  R   &  --   \\
51 * &  B   &      &  --  &  B   &   111 &  B   &  R   &  --  &  B    \\
\hline
\end{tabular}
\end{center}
\end{table*}

\begin{figure}
  \centering
  \includegraphics[angle=-90, width=8cm]{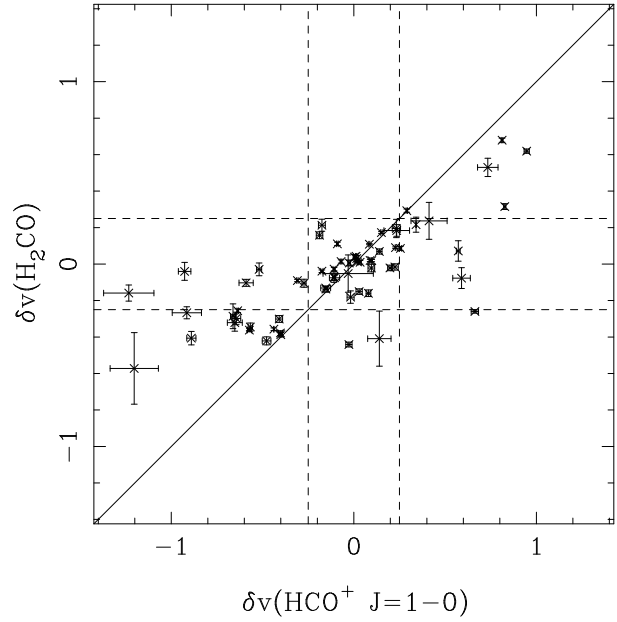}
  \caption{Comparison of the measured asymmetry in \hcoplus\ \j{1}{0}\ and 
  \htwoco. The dashed lines mark 
  line of equality is shown in solid.}
  \label{fig:compasymm}
\end{figure}

 A comparison of the asymmetry parameter measured towards all the sources
 observed in both \hcoplus\ \j{1}{0}\ and \htwoco\ is shown in
 Figure~\ref{fig:compasymm}. The asymmetry of these lines is relatively well
 correlated, but for majority of the sources the asymmetry of the \htwoco\ is
 less than half that of the \hcoplus\ perhaps suggesting that these two lines
 are probing some what different regions around the stars. This would be
 consistent with the different critical densities of these two transitions,
 $5\times10^4$ cm$^{-3}$ and $2\times10^6$ cm$^{-3}$ for the \hcoplus\ and
 \htwoco\ respectively. Interestingly the two transitions which show the least
 blue asymmetry, \hcoplus\ \j{3}{2}\ and \j{4}{3}, also have the highest
 critical densities ($3\times10^6$ cm$^{-3}$ and $9\times10^6$ cm$^{-3}$
 respectively) perhaps suggesting stronger infall in the lower density, outer
 regions of these sources. However until the molecular excitation and line
 formation in the circumstellar regions of these sources have been modelled in
 detail, this should only be regarded as speculation.

\subsection{Comparison with Other Surveys}

The level of blue excess seen in this sample of high-mass objects is
comparable to that accepted as evidence of infall in similar surveys of
low-mass star forming regions.  For example, over their whole sample
\citet{mar} find an excess of 0.25, while for several surveys using \hcoplus\ 
\j{3}{2}\ transitions, \citet{evans03} quotes excesses of 0.30 to 0.31.

In a recent survey of a different sample of young high mass sources
using different tracers \citet{wue} found an excess of 0.21 to 0.29,
again very similar to the values found here. Interestingly there is
one object in common between this survey and that of Wu \& Evans. The
source WFS~79 (IRAS~19410+2336) is listed in Wu \& Evans as
$59.78+0.06$.  While they find a velocity for this source of
$22.63\pm0.07$\kms\ and an optically linewidth of 2.36\kms, the
\ntwohplus\ transition is somewhat narrower with a width of 1.8\kms
and a slightly lower velocity of $22.52\pm0.01$\kms. Nevertheless, 
both surveys find blue asymmetric line profiles towards this source
with all three transitions observed here significantly blue
asymmetric.

\section{Infall Candidates}
\label{sec:infallc}

The observations of multipe transitions towards the sources in this survey
provides a second approach to investigate whether there is infall around the
sources. Since infall is expected to preferentially produce blue asymmetric
lines, it seems reasonable to expect that this asymmetry should be present in
different transitions observed towards the sources. In particular the presence
of multiple blue asymmetric transitions and the absence of contradictory
information, in the form of red asymmetric profiles, could provide a good
criterion for identifying sources with infall, independent of the overall
excess of blue line profiles in the sample.

Table~\ref{tab:linesum} summarises those sources which have at least
one significantly asymmetric line and from this table we can identify
those sources we believe to the best infall candidates. We do this by
requiring an infall candidate to have at least one blue asymmetric
transition and no red asymmtric line profiles.  This criterion, which
is the same as that used by \citet{mar} produces 22 promising infall
candidates. The majority of these candidates, 14 sources, in fact have
two or more blue asymmetric profiles and no red profiles.

\begin{figure}
  \centering
  \includegraphics[angle=0, width=16cm]{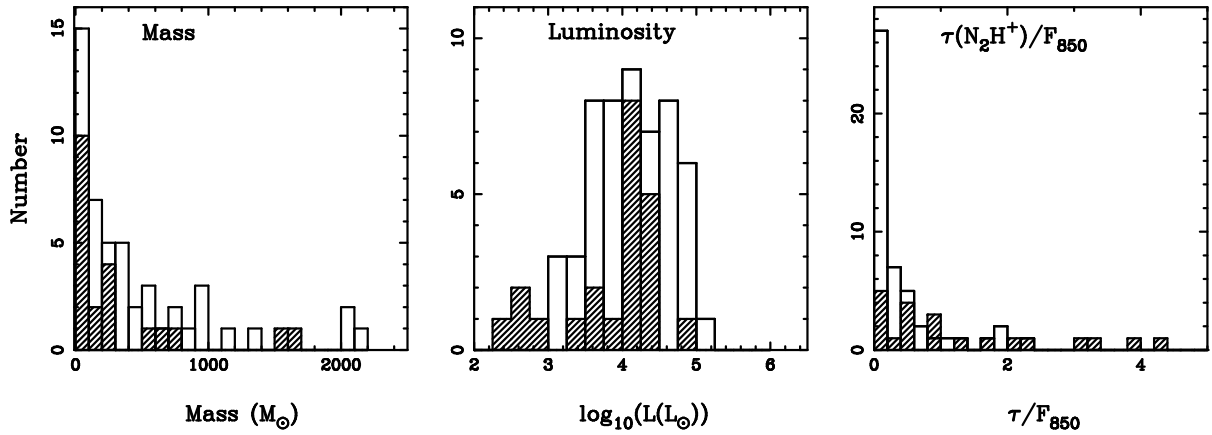}
  \caption{Comparison of the properties of the infall candidates (shaded)
    and the other sources in the sample (outline).  The left panel shows the
    distrbution of mass derived from the 850\microns\ observations of
    \citet{wfs1} for the infall candidates and the non-infall sources. The
    centre panel similiarly compares the distribution of the IRAS luminosity,
    assuming the near kinematic distances for the sources where the distance
    is ambiguous \citep{tk1}. The right panel shows the distribution of the
    ratio of the \ntwohplus\ optical depth to the 850\microns\ flux as
    discussed in Section~\ref{sec:infallc}.}
  \label{fig:compsrcs}
\end{figure}

The 22 infall candidates range in luminosity from $10^{2.4}$\Lsun\ up to
$10^{4.8}$\Lsun\ with a median luminosity of $10^{4.0}$\Lsun\ and span a range
of distances from 0.6kpc to 10.6kpc but only two sources are at distances
greater than 5.5 kpc, assuming near kinematic distances for sources whose
distance ambiguity has not been resolved.  At these distances the JCMT and
IRAM beams are sampling regions 0.02pc to 0.4pc (0.6 to 12$\times10^{17}$cm)
in radius around the sources.  The submillimetre derived mass of the cores
around the infall candidates range from several to over a thousand solar
masses with a median value of 144\Msun.

A comparison of IRAS derived luminosity of the sources \citep{tk1} and the
mass derived from the 850\microns\ observations \citep{wfs1} for the infall
candidates and the other sources in the sample is shown in
Figure~\ref{fig:compsrcs}.  Applying a two sample Kolmogorov-Smirnov (K-S)
test shows that there is no evidence for a difference in either mass or
luminosity between the two groups of objects.  Similarly K-S tests show there
is no significant difference in the \ntwohplus\ linewidths, 850\microns\ 
column density, mass to luminosity ratio, 850\microns\ to 450\microns\ 
spectral index or cold component dust temperature \citep{tk1} between the
infall candidates and other sources. Inspection of the radio continuum
emission and maser properties \citep{tk1} also shows that the infall
candidates do not have any different range of properties to the other sources
in the sample.

However, the K-S test does indicate that there are differences (at a
confidence level of about 98\%) between the infall candidates and the
remainder of the sample in terms of the optical depth of the \ntwohplus\ and
the 850\microns\ peak flux and, at a somewhat lower significance level, the
850\microns\ integrated flux.  The distributions of these quantities suggest
that the infall candidates have higher \ntwohplus\ optical depth and less dust
continuum emission than the other sources.  Figure~\ref{fig:compsrcs} shows
the distributions of the ratio of the \ntwohplus\ optical depth to
850\microns\ peak flux, which shows a similiar difference between the infall
and non-infall sources with the infall candidates having a higher ratio of
\ntwohplus\ to dust emission.  Although the statistical significance of this
difference is not conclusive, if real it has interesting implications about
the state of these sources.  \ntwohplus\ is primarily destroyed by reacting
with CO \citep{h75}, leading to high \ntwohplus\ abundances in regions in
which CO is depleted \cite{be01}.  Larger CO depletion towards the infall
candidates might suggest that these sources are less evolved having not (yet)
been as extensively heated by their embedded sources as the non-infall
sources. A direct test of this will be to measure the CO depletion towards
these sources.

\subsection{Infall Properties}

\begin{figure}
  \centering
  \includegraphics[angle=-90, width=4cm]{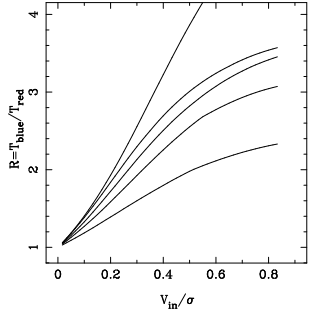}
  \caption{Line peak intensity ratio 
    $R=T_{\rm blue}/T_{\rm red}$ as a function of infall velocity in
    units of the velocity dispersion, $V_{\textrm{in}}/\sigma$. The line
    profiles were calculated for five different models using the two
    layer model of \citet{mm}. }
  \label{fig:mm}
\end{figure}

If further observations of these candidate sources supports their
identification as having infall, then detailed modelling of the temperature,
density, velocity and indeed chemical structure of the sources will be needed
to derive the properties of the infall. However an order of magnitude estimate
of the mass accretion rate can be derived from applying the two layer model of
\citet{mm}.  In an infalling region the strength of the blue-red asymmetry of
a line, as measured by $R$ the ratio of the intensity of the blue line peak
compared to the red line peak, depends strongly on the infall velocity.
Figure~\ref{fig:mm} shows this intensity ratio as a function of the infall
velocity ($V_{\rm in}$) in units of the velocity dispersion ($\sigma$)
calculated using the Myers et al.  two layer model.

For lines with small asymmetries, $R=T_{\rm blue}/T_{\rm red}\simlt1.5$, the
infall velocity is relatively tightly constrained. For $R=1.5$ the infall
velocity is between $\sim0.1$ and $\sim0.25$ times the velocity dispersion.
For more asymmetric lines the infall velocity exceeds 0.25 times the velocity
dispersion and can become comparable to, or greater, than the velocity
dispersion.  For the sources and transitions in this survey with blue
asymmetric lines which show double peak profiles, the value of $R$ is always
in excess of 1.2 with values ranging up to 3, implying infall velocities from
0.15 to $>1$ times the velocity dispersion. There are also clearly blue
asymmetric line profiles for which a red peak can not be identifed suggesting
even larger infall velocities. Adopting the \ntwohplus\ linewidth as a measure
of the velocity dispersion in the circumstellar material this translates into
infall velocities between $\sim0.1$ \kms\ and 1 \kms\ with a typical value of
$\sim0.2$ \kms.  Adopting a size for the infall region similar to the
submillimeter maps of WFS, $9\times10^{17}$cm and a number density of
$5\times10^4$\cc\ this range of infall velocities implies mass infall rates of
between $2\times10^{-4}$\Msun/yr and $10^{-3}$\Msun/yr.

This range of infall rates is consistent that values inferred from
observations of outflows from these sources \citep{b1}, although those rates
refer to the region where the stellar winds are generated, much closer the
central source that the observations presented here.  Interestingly the lower
end of the range of infall rates is also similar to the rates derived for some
low-mass sources. For example,  the low lumniosity source NGC 1333 IRAS4A
has an infall rate of $1.1\times10^{-4}$\Msun/yr \citep{dif}.

\section{Summary}

A survey of up to six transitions of \hcoplus\, \hthirteencoplus\ 
\htwoco\ and \ntwohplus\ has been carried out towards a sample of 77
submillimetre sources associated with a well selected sample of IRAS
sources believed to be high mass protostellar objects.

Emission from \ntwohplus\ was detected towards all the sources except for a
few cases where the observation was significantly offset from the peak of the
submillimetre emission.  The strength of the \ntwohplus\ towards the majority
of sources and its weakness towards these offset positions suggests that the
\ntwohplus\ and dust continuum emission are tracing similar material.  The
\ntwohplus\ is mostly optically thin and typically well modelled by a single
Gaussian velocity component.  The \ntwohplus\ velocity and linewidth are
typically in good agreement with those measured for the same source in
\hthirteencoplus, which is also believed to be optically thin and is also well
modelled by a single Gaussian component.

Using the \ntwohplus\ velocity as a measure of the systemic velocity of the
circumstellar material associated with the forming stars in these sources, the
lines of \hcoplus\ and \htwoco\ show a range of profile shapes from near
Gaussian, centred at the systemic velocity or offset, to double peaked.
Analysis of the asymmetry of the line profiles shows that in \hcoplus\ 
\j{1}{0}\ and \htwoco\ there is a statistically significant excess of blue
asymmetric line profiles compared to red asymmetric profiles. This excess is
larger if the sample is restricted to exclude the most distant sources. Since
the emission from infalling material around a central heating source is known
to produce systematically blueshifted line profiles, this excess is
interpreted as statistical evidence that the material around these sources is
infalling, as has been argued for similar samples of nearby low mass
protostars.

From the observations of multiple lines, we identify a sample of 22
strong infall candidates none of which show any red asymmetric line
profiles and have at least one blue asymmetric profile. Confirmation
of the presence of infalling material around these sources requires
further observations, in particular maps of the sources to study the
spatial distribution of the blue asymmetric profiles and higher
angular resolution observations to better probe the inner material
close to the central sources, together with detailed modelling of the
sources.  Nevertheless, the blue asymmetric line profiles together
with density and size estimates from submillimetre maps of the sources
allow estimates of the range of infall velocity which is found to be
between 0.1 \kms\ and 1 \kms\, and the mass infall rate which is
estimated to be between $2\times10^{-4}$\Msun/yr and
$10^{-3}$\Msun/yr.  Even the smallest estimate is consistent
with the mass infall rates needed to form a 10\Msun\ in $\sim10^5$
years \citep{mt02}.


\begin{acknowledgements}
  Part of this work was supported by a PPARC grant to the UMIST
  Astrophysics Grant and a PPARC studentship to SJW. We would like to
  thank the staff at both the JCMT and IRAM 30m for their assistance
  in aquiring the observations presented here.
\end{acknowledgements}

\end{document}